\documentclass[a4paper,12pt]{article}

\usepackage{amsmath}
\usepackage{amsfonts}
\usepackage{graphicx}
\usepackage[nosort]{cite}

\numberwithin{equation}{section}

\def\beq#1\eeq{\begin{equation}#1\end{equation}}
\def\bes#1\ees{\begin{equation}\begin{split}#1\end{split}\end{equation}}
\def\bea#1\eea{\begin{align}#1\end{align}}

\newcommand{\abs}[1]{\lvert #1 \rvert}

\newcommand{\Z}{\mathbb{Z}}
\newcommand{\Q}{\mathbb{Q}}

\newcommand{\g}{\mathfrak{g}}
\newcommand{\h}{\mathfrak{h}}

\newcommand{\bra}[1]{\langle #1|}
\newcommand{\ket}[1]{|#1\rangle}
\newcommand{\dbra}[1]{\langle\!\langle #1|}
\newcommand{\dket}[1]{|#1 \rangle\!\rangle}

\newcommand{\auto}[1]{G(#1)}
\newcommand{\fusion}[4][\mathcal{N}]{{#1}_{#2 #3}{}^{#4}}

\newcommand{\A}{\mathcal{A}}
\newcommand{\I}{\mathcal{I}}
\newcommand{\HH}{\mathcal{H}}
\newcommand{\E}{\mathcal{E}}
\newcommand{\V}{\mathcal{V}}

\newcommand{\G}{G_\mathrm{sc}}          
\newcommand{\Gid}{G_\mathrm{id}}        
\newcommand{\stab}[1]{\mathcal{S}(#1)}  
\newcommand{\stabhat}{\widehat{\mathcal{S}}(\V)}  

\DeclareMathOperator*{\dirsum}{\oplus}
\DeclareMathOperator{\trace}{Tr}

\newcommand{\DD}{\mathcal{D}}
\newcommand{\NN}{\mathcal{N}}
\newcommand{\TT}{\mathcal{T}}

\begin{document}
\baselineskip=6mm

\begin{titlepage}
\nopagebreak
\vskip 5mm
\begin{flushright}
hep-th/0301040\\
TU-678\\
UT-Komaba/2003-1
\end{flushright}

\vskip 15mm
\begin{center}
\baselineskip=10mm
{\LARGE
\textbf{%
Twisted boundary states in $c=1$ coset conformal field theories}}
\end{center}
\begin{center}
\vskip 12mm
Hiroshi \textsc{Ishikawa}$^1$ and Atsushi \textsc{Yamaguchi}$^2$
\vskip 5mm
\textsl{%
$^1$Department of Physics, Tohoku University \\
Sendai 980-8578, JAPAN\\\vspace{10pt}
$^2$Institute of Physics, University of Tokyo\\
Komaba 153-8902, JAPAN\\
}
\vspace{10pt}
\texttt{%
\footnotesize
$^1$ishikawa@tuhep.phys.tohoku.ac.jp \\
$^2$yamaguti@hep1.c.u-tokyo.ac.jp
}
\end{center}

\vspace{30pt}
\begin{abstract}
\baselineskip=6mm
We study the mutual consistency of twisted boundary conditions
in the coset conformal field theory $G/H$.
We calculate the overlap of the twisted boundary states of $G/H$
with the untwisted ones, and show that the twisted boundary states
are consistently defined in the diagonal modular invariant.
The overlap of the twisted boundary states is expressed by
the branching functions of a twisted affine Lie algebra.
As a check of our argument, we study the diagonal coset theory
$so(2n)_1 \oplus so(2n)_1/so(2n)_2$, which is equivalent
with the orbifold $S^1/\Z_2$.
We construct the boundary states twisted by the automorphisms
of the unextended Dynkin diagram of $so(2n)$, and
show their mutual consistency by identifying their counterpart
in the orbifold.
For the triality of $so(8)$, the twisted states of the coset
theory correspond to neither the Neumann nor the Dirichlet
boundary states of the orbifold and yield the conformal boundary states
that preserve only the Virasoro algebra.
\end{abstract}

\vfill
\end{titlepage}

\section{Introduction}

The nature of boundary conditions in rational conformal field theories
(RCFTs) has been now fairly well understood
\cite{Cardy,CardyLewellen,Lewellen,Pradisi,FS2,Runkel,FS,BFS,BPPZ,FFFS}.
There are a set of conditions that
every consistent set of the boundary states
should satisfy, which is called the sewing relations
\cite{CardyLewellen,Lewellen}.
The Cardy condition \cite{Cardy} is one of the sewing relations
and expresses the consistency of the annulus amplitudes without
insertion of vertex operators.
It is now recognized that
solving the Cardy condition is equivalent to finding
the non-negative integer valued matrix representation (NIM-rep)
of the fusion algebra\cite{BPPZ,Gannon} under the assumption
of the completeness \cite{Pradisi}.
The classification problem of the possible boundary conditions
in the given RCFT is therefore related with those of NIM-reps of the fusion
algebra.

Since the fusion coefficients themselves form a representation of the fusion
algebra, we always have at least one NIM-rep for each fusion algebra,
which is realized by the fusion matrices and called the regular one.
Therefore the point is whether we have a non-trivial NIM-rep other than
the regular one.
If we do not require an integer-valued representation,
we have several one-dimensional ones realized by the (generalized)
quantum dimensions, which in general take values in irrational
numbers.
However the requirement for a representation to be NIM is so severe
that one finds a non-trivial NIM-rep
in only the limited cases.
On the other hand, it happens that
the number of possible NIM-reps is greater than one
and that we have several solutions for one particular fusion algebra.
Although these different NIM-reps are in general considered to be
corresponding to different modular invariants,
it is possible that two or more NIM-reps coexist in the
same modular invariant and are compatible with each other.

This issue of mutual consistency of different NIM-reps
has been studied in \cite{BFS,GG} (see also \cite{Ishikawa})
for the case of the WZW models.
In the WZW model based on a group $G$, 
the chiral algebra is an (untwisted) affine Lie algebra $\g$,
and the regular NIM-rep describes the boundary conditions
preserving $\g$.
If the horizontal subalgebra of $\g$ has an automorphism $\omega$,
we can twist $\g$ by $\omega$ to obtain a non-trivial 
boundary condition in the WZW model, and
one can find a non-trivial NIM-rep corresponding to the twisted
boundary condition. Eventually we obtain two types of
NIM-reps in one theory:
the regular and the twisted ones.
The mutual consistency of these two has been shown 
by calculating the annulus amplitude between two types of
the boundary states.
The twisted affine Lie algebra associated with $\g$ and
$\omega$ plays an essential role in the calculation.

In this paper, 
we extend the analysis of the mutual consistency of different NIM-reps
for the WZW models
to an important class of CFTs, coset conformal field theories \cite{GKO}.
The boundary states in coset theories has been studied by several groups
\cite{MMS,Gawedzki,ES,Fredenhagen,Ishikawa,Kubota,Quella,%
Fredenhagen2,Walton,IT}.
In particular, it has been shown that one can construct a NIM-rep
of the $G/H$ theory starting from those of $G$ and $H$ \cite{Ishikawa,IT}.
Hence a non-trivial NIM-rep of $G/H$ follows from the twisted NIM-reps
of $G$ and $H$, if $G$ (or $H$) is equipped with 
an automorphism \cite{MMS,Ishikawa}.
We calculate the overlap of the corresponding twisted Cardy states of $G/H$
with the untwisted ones, and show that the twisted states
are consistently defined in the diagonal modular invariant.
The overlap of the twisted boundary states is given by
the branching functions of the twisted affine Lie algebras associated
with the automorphism used for twisting the boundary condition.
As a check of our argument, we study in detail the diagonal coset theory
$so(2n)_1 \oplus so(2n)_1/so(2n)_2$, which is equivalent
with the orbifold $S^1/\Z_2$.
We construct the boundary states twisted by the automorphisms
of the unextended Dynkin diagram of $so(2n)$, and
show their mutual consistency by identifying their counterpart
in the orbifold.
For the order-2 automorphism of $so(2n)$, the resulting states
as well as the regular ones are realized by the usual Neumann
and Dirichlet boundary states of the orbifold.
For the triality of $so(8)$, however, the twisted states of the coset
theory correspond to neither the Neumann nor the Dirichlet
states. Rather they yield the conformal boundary states
that preserve only the Virasoro algebra.
The conformal boundary states have been constructed for the case of $S^1$
\cite{GRW,GR,Janik,Tseng}.
Our twisted states provide their generalization to $S^1/\Z_2$. 

The organization of this paper is as follows.
We review some basic facts about the construction of NIM-reps in coset CFTs
in the next section.
In Section~\ref{sec:compatibility}, we argue the mutual consistency
of different NIM-reps in coset theories.
In particular, we show that the twisted Cardy states of $G/H$ are labeled
by the branching functions of the associated twisted affine Lie algebras
and have well-defined overlaps with the regular states.
In Section~\ref{sec:diagonal}, we apply our method
to the diagonal coset theory $so(2n)_1 \oplus so(2n)_1/so(2n)_2$
to obtain the Cardy states twisted by the automorphisms of $so(2n)$.
We treat three types of automorphisms: 
the order-$2$ automorphism $\omega_2$ of $so(2n)$,
the order-$3$ automorphism $\omega_3$ of $so(8)$ and
the permutation automorphism $\pi$ of $so(2n) \oplus so(2n)$.
For $\omega_2$ and $\omega_3$, we show that there exist non-trivial
NIM-reps and give them in the explicit form.
For the permutation automorphism $\pi$, however,
the resulting NIM-rep is shown to yield nothing new. 
In Section~\ref{sec:relation}, using the equivalence of the $so(2n)$
coset theory with the orbifold $S^1/\Z_2$, we map the twisted states
in the coset theory to the boundary states of the orbifold, in which
the mutual consistency of the boundary states is well understood.
We give some technical details in the Appendices.
The derivation of the conformal boundary states of $S^1/\Z_2$ is reported
in Appendix~\ref{sec:conformal}.

\section{Boundary states in coset theories}

In this section, we review the construction of
the Cardy states in coset theories developed in \cite{Ishikawa,IT}.

\subsection{Conventions}

Let $\A$ be the chiral algebra of RCFT,
which is the Virasoro algebra or an extension thereof.
We denote by $\I$ the set of
all the possible irreducible representations of $\A$,
and by $0 \in \I$ the vacuum representation.
For the WZW model based on the affine Lie algebra $\g$ at level $k$,
$\I$ is given by the set $P_+^k(\g)$ of all the integrable highest-weight
representations of $\g$ at level $k$.
Let $\HH_\mu$ be the representation space for $\mu \in \I$.
Then the character of the representation $\mu$ is written as
\beq
  \chi_\mu(\tau) = \trace_{\HH_\mu} q^{L_0 - \frac{c}{24}} ,
\eeq
where $q = e^{2\pi i \tau}$ and
$c$ is the central charge of the theory.
The characters transform among themselves under the modular transformation
$\tau \rightarrow -1/\tau$,
\beq
 \chi_{\mu}(-1/\tau)= \sum_{\nu \in \I}
                      S_{\mu\nu}\chi_{\nu}(\tau).
\label{eq:Smatrix}
\eeq
The modular transformation matrix $S$ is a symmetric unitary matrix.

The elements of $\I$ forms the fusion algebra
$(\mu) \times (\nu) =
\sum_{\rho \in \I} \fusion{\mu}{\nu}{\rho} (\rho)$.
Here $\fusion{\mu}{\nu}{\rho}$ are non-negative integers called the fusion
coefficients.
It is often useful to introduce the matrices
\beq
  (N_\mu)_\nu{}^\rho = \fusion{\mu}{\nu}{\rho} .
\label{eq:fusionmat}
\eeq
One can show that $N_\mu$  satisfy the fusion algebra
\beq
  N_\mu N_\nu = \sum_{\rho \in \I} \fusion{\mu}{\nu}{\rho} N_\rho .
\label{eq:Nfusion}
\eeq
The Verlinde formula \cite{Verlinde} relates the fusion coefficients
$\fusion{\mu}{\nu}{\rho}$ with the modular transformation
matrix $S$,
\beq
  {\mathcal{N}_{\mu \nu}}^{\rho}
     = \sum_{\lambda \in \I}
       \frac{S_{\mu \lambda}S_{\nu \lambda}
            \overline{S_{\rho \lambda}}}
            {S_{0\lambda}}
     = \sum_{\lambda \in \I}
            S_{\nu \lambda}\gamma^{(\mu)}_{\lambda}
            \overline{S_{\rho \lambda}},
\label{eq:Verlinde}
\eeq
where $\displaystyle \gamma^{(\mu)}_\lambda$
is the generalized quantum dimension
\beq
 \gamma^{(\mu)}_{\lambda}
             =\frac{S_{\mu \lambda}}{S_{0 \lambda}}.
\label{eq:qmdim}
\eeq
In terms of the matrices $N_\mu$,
the Verlinde formula can be written in the form
\beq
  {N}_{\mu} = S \gamma^{(\mu)} S^{\dagger} , \quad
  \gamma^{(\mu)} = \mathrm{diag}(\gamma^{(\mu)}_\nu)_{\nu \in \I} .
\label{eq:Verlindemat}
\eeq
The modular transformation matrix $S$ diagonalizes
the fusion matrices $N_\mu$.
The generalized quantum dimension
$\{\gamma^{(\mu)}_\lambda |\, \lambda \, \text{fixed}\}$ is therefore
a one-dimensional representation of the fusion algebra
\beq
  \gamma^{(\mu)}_\lambda \gamma^{(\nu)}_\lambda =
  \sum_{\rho \in \I} \fusion{\mu}{\nu}{\rho} \gamma^{(\rho)}_\lambda .
\label{eq:qmdimfusion}
\eeq

A simple current $J$ of a RCFT is an element in $\I$ whose
fusion with the elements in $\I$ induces
a permutation $\mu \mapsto J \mu$ of $\I$
\cite{SY,Intriligator},
\beq
  (J) \times (\mu) = (J \mu) .
\eeq
The set of all the simple currents forms an abelian group
which we denote by $\G$,
\beq
  \G = \{J \in \I \, | \,J :\,\text{simple current}\}.
\eeq
$\G$ is a multiplicative group in the fusion algebra.
The group multiplication is defined by the fusion.

A simple current $J$ acts on the modular $S$ matrix as follows,
\beq
  S_{J \lambda\, \mu} = S_{\lambda \mu} b_\mu(J),
\label{eq:sdual}
\eeq
where $b_\mu(J)$ is the generalized quantum dimension of $J$,
\footnote{%
We write $J0$ as $J$
since the action of $J$ on $0$ yields $J$ itself.
}
\beq
  b_\mu(J)  = \frac{S_{J \mu}}{S_{0 \mu}} = \gamma^{(J)}_\mu .
\label{eq:bandqmdim}
\eeq
In the matrix notation, the above relation can be
written as
\beq
  N_J S = S \, b(J) ,
\label{eq:sdualmat}
\eeq
where
$(N_J)_\mu{}^\nu = \fusion{J}{\mu}{\nu} = \delta_{J\!\mu\,\nu}$
and $b(J) = \mathrm{diag}(b_\mu(J))_{\mu \in \I}$.
Since $\displaystyle (\lambda) \mapsto \gamma^{(\lambda)}_\mu$
is a one-dimensional representation of the fusion algebra,
$J \mapsto b_\mu(J)$ is a one-dimensional representation of $\G$,
\beq
  b_\mu(J J') = b_\mu(J) b_\mu(J') , \quad J, J' \in \G .
\eeq
Therefore $b_\mu(J)$ is of the form
\beq
  b_\mu(J) = e^{2\pi i Q_\mu(J)} , \quad Q_\mu(J) \in \Q ,
  \label{eq:bQ}
\eeq
since any one-dimensional representation of a finite group takes
values in roots of unity.
The phase $Q_\mu(J)$ is called the monodromy charge.

\subsection{$G/H$ theories}

The $G/H$ theory \cite{GKO} is based on an embedding of
the affine Lie algebra $\h$ into $\g$.
A representation $\mu$ of $\g$ is decomposed by
representations $\nu$ of $\h$ as follows,
\beq
 \HH_\mu = \dirsum_{\nu} \HH_{(\mu;\nu)} \otimes \HH_\nu .
\label{eq:cosetbranching}
\eeq
The primary fields of the $G/H$ theory are labeled by $(\mu;\nu)$
and the corresponding character is given by the branching functions.
We denote by $\hat{\I}$ the set of all the possible primaries
in the $G/H$ theory,
\bes
  \hat{\I} = \{(\mu; \nu) | \,
       &\mu \in \I^G, \nu \in \I^H, \\
       &b^G_\mu(J) = b^H_\nu(J'),
        (J \mu; J' \nu) = (\mu; \nu), \forall (J, J') \in \Gid \} .
\label{eq:cosetprimary}
\ees
Here $\I^G$ $(\I^H)$ is the set of all the primaries in the $G$ $(H)$ theory.
$\Gid \subset \G^G \times \G^H$
is the group of the identification currents
corresponding to the common center of $G$ and $H$.
$b^G_\mu(J)$ ($b^H_\nu(J')$) expresses  the monodromy charge of
the $G$ ($H$) theory.
The condition $b^G_\mu(J) = b^H_\nu(J')$ is the selection rule
of the branching of representations,
while the relation $(J \mu; J' \nu) = (\mu; \nu)$ is
the field identification.

In this paper, we restrict ourselves to the case that
all the identification orbit have the same length $N_0 = \abs{\Gid}$,
\beq
  N_0 = \abs{\{(J \mu, J' \nu) \in \I^G \otimes \I^H |\,
               (J, J') \in \Gid \}}
      = \abs{\Gid} .
  \label{eq:orbitlength}
\eeq
In other words, there is no fixed point in the field identification
\cite{FSS2}
\beq
  (J \mu, J' \nu) \neq (\mu, \nu), \quad
  \forall (\mu, \nu) \in \I^G \otimes \I^H, \quad
  \forall (J, J') \neq 1 \in \Gid .
\eeq
The character of the coset theory is the branching
function $\chi_{(\mu;\nu)}$ of the algebra
embedding $h\subset g$.
From the branching rule \eqref{eq:cosetbranching}, we find
\beq
  \chi_{\mu}^G = \sum_{\nu} \chi_{(\mu;\nu)} \chi_{\nu}^H .
\eeq
The modular transformation $S$-matrix of the $G/H$ theory reads
\beq
  \hat{S}_{(\mu;\nu)(\mu';\nu')}
      = N_0 S_{\mu \mu'}^G \overline{{S_{\nu \nu'}^H}}
      = N_0 S_{\mu \mu'}^G S^H_{\nu \bar{\nu'}},
\label{eq:cosetSmatrix}
\eeq
where $N_0$ is the length of the identification orbit
\eqref{eq:orbitlength} and $S^G$ ($S^H$) is the $S$-matrix of
the $G$ ($H$) theory.
$\bar{\nu'}$ is the representation conjugate to $\nu'$.

The fusion coefficients of the coset theory can be calculated
via the Verlinde formula \eqref{eq:Verlinde},
\bes
  \hat{\mathcal{N}}_{(\mu; \nu)(\mu'; \nu')}{}^{(\mu''; \nu'')}
  &= \sum_{(\rho;\sigma) \in \hat{\I}}
    \frac{%
          \hat{S}_{(\mu; \nu)(\rho; \sigma)}
          \hat{S}_{(\mu'; \nu')(\rho; \sigma)}
          \overline{\hat{S}_{(\mu''; \nu'')(\rho; \sigma)}}}
          {\hat{S}_{(0; 0)(\rho; \sigma)}}  \\
  &= \frac{1}{N_0} \sum_{(\rho, \sigma) \in \I^G \otimes \I^H}
         \frac{1}{N_0} \sum_{(J, J') \in \Gid}
                        b^G_{\rho}(J) b^H_{\sigma}(J')^{-1} \\
         & \quad\quad\quad\quad\quad\quad\quad\quad\quad
         \times N_0^2
         \frac{%
          S^G_{\mu\rho} S^G_{\mu' \rho} \overline{S^G_{\mu'' \rho}}}
         {S^G_{0 \rho}}
         \frac{%
          S^H_{\nu \bar{\sigma}} S^H_{\nu' \bar{\sigma}}
               \overline{S^H_{\nu'' \bar{\sigma}}}}
         {S^H_{0 \bar{\sigma}}} \\
  &= \sum_{(J, J') \in \Gid}
     \mathcal{N}^G_{\mu\, J\!\mu'}{}^{\mu''}
     \mathcal{N}^H_{\nu\, J'\!\nu'}{}^{\nu''} .
\label{eq:coset_fusion}
\ees
Here we used our assumption of no fixed points to rewrite the sum
\beq
  \sum_{(\rho; \sigma) \in \hat{\I}} \rightarrow \quad
     \frac{1}{N_0} \sum_{(\rho, \sigma) \in \I^G \otimes \I^H}
     \frac{1}{N_0} \sum_{(J, J') \in \Gid}
                        b^G_{\rho}(J) b^H_{\sigma}(J')^{-1} .
\eeq
The projection operator introduced above takes account of the selection
rule.

\subsection{NIM-reps in RCFTs}
In the bulk theory, we have a pair of algebras $\A$ and $\tilde{\A}$,
which correspond to the holomorphic and the anti-holomorphic
sectors, respectively.
The space $\HH$ of the states in the bulk theory can therefore
be decomposed into the representations of $\A \otimes \tilde{\A}$ as
\beq
 \HH = \bigoplus_{\mu, \nu \in \I}
 M_{\mu \nu} \, \HH_{\mu} \otimes \widetilde{\HH}_{\bar{\nu}} \quad
 (M_{\mu \nu} \in \Z_{\ge 0} ) .
\label{eq:spec}
\eeq
The modular invariant partition function associated with this spectrum reads
\beq
  Z(\tau) = \sum_{\mu, \nu \in \I}
  M_{\mu \nu} \, \chi_\mu(\tau)  \overline{\chi_\nu(\tau)} .
\label{eq:partition}
\eeq
In this paper, we restrict ourselves to the case of the diagonal
invariant $M_{\mu \nu} = \delta_{\mu \nu}$.

In the presence of boundaries, $\A$ and $\tilde{\A}$ are related with
each other via appropriate boundary conditions.
In the closed string channel,
we can describe boundary conditions in terms of boundary states,
which satisfy the following
gluing condition
\beq
   (\omega(W_n) - (-1)^h \widetilde{W}_{-n}) \ket{\alpha}_\omega =0 .
\label{eq:glue}
\eeq
Here $W$ and $\widetilde{W}$ are the currents of
$\A$ and $\tilde{\A}$, respectively.
$h$ is the conformal dimension of the currents and
$\omega$ is an automorphism of the chiral algebra $\A$.
$\omega$ induces an isomorphism $\mu \mapsto \omega \mu$
of the representations $\mu \in \I$ of the chiral algebra $\A$.
In order to keep the conformal invariance, $\omega$ has to fix
the generators $L_n$ of the Virasoro algebra,
\beq
  \omega(L_n) = L_n .
\eeq
In the following,
a boundary condition for $\omega \neq 1$ will be called
the \textit{twisted} type,
while $\omega = 1$ the \textit{untwisted} type.

We denote by $\V$ the set labeling the boundary states.
A generic boundary state $\ket{\alpha} (\alpha \in \V)$
satisfying the boundary condition \eqref{eq:glue} is
a linear combination of the Ishibashi states \cite{Ishibashi}
\beq
  \ket{\alpha}_\omega = \sum_{\mu \in \E}
               {\psi_{\alpha}}^{\mu} \frac{1}{\sqrt{S_{0\mu}}}
               \dket{\mu}_\omega  \quad \quad
               (\alpha \in \V) .
\label{eq:boundary}
\eeq
Here $\E$ is a set labeling the Ishibashi states
compatible with the gluing automorphism $\omega$ and the modular
invariant $Z$.
For the charge-conjugation modular invariant, $\E$ is given by the set
$\I^\omega$ of all the representations fixed by $\omega$,
\beq
  \I^\omega = \{ \mu \in \I \,| \, \omega \mu = \mu \} .
\label{eq:specomega}
\eeq
We normalize the Ishibashi states as follows,
\beq
  {}_\omega\dbra{\mu} q^{H_c} \dket{\nu}_\omega =
  \delta_{\mu\nu}\chi_\mu(\tau) \quad  (\mu, \nu \in \I^\omega) ,
\label{eq:Ishibashinorm}
\eeq
where
$H_c = \frac{1}{2}(L_0 + \tilde{L}_0 - \frac{c}{12})$
is the closed string Hamiltonian.

The coefficients ${\psi_{\alpha}}^{\mu}$ together with the sets $\E$
and $\V$ should be chosen appropriately
for the mutual consistency of the boundary states \cite{Cardy,BPPZ}.
Consider the annulus amplitude between two
boundary states $\ket{\alpha}$ and $\ket{\beta}$,
\beq
 Z_{\alpha \beta}=\bra{\beta}\tilde{q}^{H_c}\ket{\alpha}
                 = \sum_{\nu \in \E,
                         \mu \in \I}
                   {\psi_{\alpha}}^{\nu}
                   \frac{S_{\mu \nu}}
                        {S_{0 \nu}}
                   \overline{{\psi_{\beta}}^{\nu}}\,
                   \chi_{\mu}(\tau)
  = \sum_{\mu \in \I} n_{\mu \alpha}{}^\beta \chi_\mu(\tau)
\quad\quad (\alpha, \beta \in \V) .
\eeq
Here we denote by $n_{\mu \alpha}{}^{\beta}$ the multiplicity
of the representation $\mu \in \I$
in $Z_{\alpha \beta}$,
\beq
  {n_{\mu \alpha}}^{\beta}
     =\sum_{\nu \in \E}
     {\psi_{\alpha}}^{\nu}
                   \gamma^{(\mu)}_{\nu}
                   \overline{{\psi_{\beta}}^{\nu}} .
\label{eq:Cardy}
\eeq
In the matrix notation, this can be written as
\beq
  n_\mu = \psi \gamma^{(\mu)} \psi^\dagger, \quad
  {(n_\mu)_{\alpha}}^{\beta}={n_{\mu \alpha}}^{\beta}, \quad
  {(\psi)_{\alpha}}^{\mu}={\psi_{\alpha}}^{\mu} .
\label{eq:Cardymat}
\eeq
Clearly, $n_\mu$ are matrices with
non-negative integer entries for consistent boundary states.
In particular, $n_0 = 1$ for the uniqueness of the vacuum
in the open string channel.
This condition for the consistency of the annulus amplitude
is called the Cardy condition and the boundary states satisfying
the Cardy condition are called the Cardy states.
Under the assumption of the completeness
$\abs{\V} = \abs{\E}$ \cite{Pradisi},
the Cardy condition implies that $\psi$ is a unitary matrix.
The matrices $n_\mu$ are therefore related with the generalized quantum
dimensions via a similarity transformation \eqref{eq:Cardymat}
and satisfy the fusion algebra
\beq
  n_\mu n_\nu = \sum_{\rho \in \I} \fusion{\mu}{\nu}{\rho} n_\rho , \quad
  n_\mu^T = n_{\bar{\mu}} .
\eeq
Since $n_{\mu \alpha}{}^\beta$ take values in $\Z_{\ge 0}$,
$\{n_\mu \}$ forms a non-negative integer matrix representation
(NIM-rep) of the fusion algebra \cite{BPPZ}.
For each set of the mutually consistent boundary states,
we have a NIM-rep $\{n_\mu | \, \mu \in \I \}$ of the fusion algebra.
The simplest example is the regular NIM-rep
\beq
  \psi = S   \quad (\E=\V=\I), \quad n_{\mu}=N_{\mu},
\label{eq:Atype}
\eeq
which corresponds to the trivial gluing automorphism $\omega = 1$ in
the diagonal modular invariant
and exists in any RCFT.
We note that
there are many `unphysical' NIM-reps that do not correspond
to any modular invariant \cite{Gannon}.
This fact shows that the Cardy condition is not a sufficient
but a necessary condition for consistency.

\medskip
The similarity between $S_{\mu \nu}$ and $\psi_{\alpha}{}^{\mu}$
\eqref{eq:Verlindemat}\eqref{eq:Cardymat} suggests that
the simple currents act also on a diagonalization matrix $\psi$.
As is discussed in \cite{Gannon,Ishikawa,GG,IT},
we actually have two types of actions of
the simple currents on $\psi$,
\begin{subequations}
\label{eq:sconNIM}
\bea
  \psi_{J \alpha}{}^\mu &= \psi_\alpha{}^\mu \, b_\mu(J), &
  n_J \psi &= \psi \, b(J) , &
  J &\in \auto{\V} ,
  \label{eq:sconbrane} \\
  \psi_{\alpha}{}^{J\mu} &= \tilde{b}_\alpha(J) \, \psi_\alpha{}^\mu, &
  \psi N_J^T &= \tilde{b}(J) \psi , &
  J &\in \auto{\E}.
  \label{eq:sconspec}
\eea
\end{subequations}
Here $\auto{\V}$ and $\auto{\E}$ are the groups of simple currents
for $\V$ and $\E$,
\begin{subequations}
\label{eq:Gpsi}
\bea
  \auto{\V} &= \G /\stab{\V} , \\
  \auto{\E} &= \{ J\in \G \,|\, J: \, \E \rightarrow \E\} ,
\eea
\end{subequations}
where $\stab{\V}$ is the stabilizer of $\V$, \textit{i.e.},
a subgroup of $\G$ which acts trivially on $\V$,
\beq
  \stab{\V}
     =\{ J_0 \in \G \,|\, J_0 \alpha = \alpha ,\,
                         \forall \alpha \in \V \}
     =\{ J_0 \in \G \,|\, b_{\mu}(J_0)=1 ,\,
                         \forall \mu \in \E \}.
  \label{eq:stabilizer}
\eeq
$\tilde{b}_\alpha(J)$ are the counterpart of $b_\mu(J)$ and defined as
\beq
  \tilde{b}_\alpha(J)  =
  \frac{\psi_{\alpha}{}^{J}}{\psi_{\alpha}{}^{0}} .
\label{eq:btilde}
\eeq
The group $\auto{\V}$ acts on $\V$ as a permutation of the Cardy states,
and the group
$\auto{\E}$ enables us to define the ``charge'' $\tilde{b}_\alpha$
of the Cardy states,
which takes values in the dual group of $\auto{\E}$.
Clearly, for the regular NIM-rep $\psi = S$, these two groups
coincide with the simple current group $\G$ itself.

The transformation property \eqref{eq:sconbrane} follows from
the fact that $n_J$ is a permutation matrix on $\V$ \cite{Gannon,Ishikawa}.
The second one \eqref{eq:sconspec} is related \cite{Gannon}
with the following algebra \cite{Pradisi,FS2,RS,BPPZ}
\beq
  \frac{\psi_{\alpha}{}^{\mu}}{\psi_{\alpha}{}^{0}}
  \frac{\psi_{\alpha}{}^{\nu}}{\psi_{\alpha}{}^{0}} =
  \sum_{\rho \in \E} \mathcal{M}_{\mu \nu}{}^\rho
  \frac{\psi_{\alpha}{}^{\rho}}{\psi_{\alpha}{}^{0}} ,
\label{eq:Malgebra}
\eeq
which coincides in some cases with 
the algebra that appeared in the classification problem
of the bulk theory \cite{Pasquier}.
Actually, if we set
\beq
  (M_\mu)_\nu{}^\rho = \mathcal{M}_{\mu \nu}{}^\rho , \quad
  \tilde{\gamma}^{(\mu)} =
  \textrm{diag}\!
  \left(
    \frac{\psi_{\alpha}{}^{\mu}}{\psi_{\alpha}{}^{0}}
  \right)_{\alpha \in \V} ,
\eeq
the algebra \eqref{eq:Malgebra} can be written in the form
\beq
  \psi M_{\mu}^T  =  \tilde{\gamma}^{(\mu)} \psi  .
\eeq
Hence, the transformation property \eqref{eq:sconspec} follows
from eq.\eqref{eq:Malgebra} if
$M_J (J \in \Gid(\E))$ coincides with
$N_J$ restricted on $\E$.

\medskip
We are now ready to state the construction of NIM-reps
in $G/H$ theories \cite{Ishikawa,IT}.
The starting point of the construction is a pair of
NIM-reps $(\psi^G, \psi^H)$ of the $G$ and the $H$ theories
with the sets $\V^G, \E^G, \V^H$ and $\E^H$.\footnote{%
This can be generalized to the case of NIM-reps not factorizable
into the $G$ and the $H$ parts \cite{IT}.
}
Then one can make a NIM-rep of the $G/H$ theory in the following form
\beq
  \hat{\psi}_{(\alpha; \beta)}{}^{(\mu; \nu)}
  = \sqrt{\abs{\Gid(\E)} \abs{\Gid(\V)}} \,
    \psi^{G}_\alpha{}^\mu  \psi^{H}_\beta{}^{\bar{\nu}} , \quad
  (\alpha; \beta) \in \hat{\V} , \,
  (\mu; \nu) \in \hat{\E} .
\label{eq:finalpsi}
\eeq
Here, $\hat{\V}$ and $\hat{\E}$ are labels of the Cardy states and
the Ishibashi states of $G/H$, respectively, and defined as
\bea
  \hat{\V} &= \{(\alpha; \beta) \, | \,
        \alpha \in \V^G, \beta \in \V^H, \,
        \tilde{b}^G_\alpha(J) = \tilde{b}^H_\beta(J') \,
        (\forall (J, J') \in \Gid(\E)), \notag \\
&\quad\quad\quad\quad\quad\quad\quad\quad\quad\quad\quad\quad
        (J \alpha; J' \beta) = (\alpha; \beta) \,
        (\forall (J, J') \in \Gid(\V)) \} ,
\label{eq:cosetboundary} \\
  \hat{\E} &= \{(\mu; \nu) \, | \,
        \mu \in \E^G, \nu \in \E^H, \,
        b^G_\mu(J) = b^H_\nu(J')  \,
        (\forall (J, J') \in \Gid(\V)), \notag \\
&\quad\quad\quad\quad\quad\quad\quad\quad\quad\quad\quad\quad
        (J \mu; J' \nu) = (\mu; \nu) \,
        (\forall (J, J') \in \Gid(\E)) \} .
\label{eq:cosetspec}
\eea
$\Gid(\E)$ and $\Gid(\V)$ are the groups of the identification
currents for the Ishibashi states and the Cardy states, respectively,
\bea
  \Gid(\E) &=
  \Gid \cap (\auto{\E^G} \times \auto{\E^H}) ,
  \label{eq:Gidspec} \\
  \Gid(\V) &= \Gid / (\Gid \cap \stabhat) ,
\label{eq:Gidbrane}
\eea
where $\stabhat$ is the stabilizer for the action of the simple currents
on the Cardy states
\footnote{%
In \cite{IT}, $\stabhat$ is denoted by $\mathcal{S}^c(\V^c)$.
}
\beq
  \stabhat = \{(J, J') \in \G |\,
   b^G_\mu(J) = b^H_\nu(J') \,\,
                        \forall (\mu, \nu) \in \E^G \otimes \E^H \} .
\label{eq:stabhat}
\eeq
Let $n^G (n^H)$ be the NIM-rep matrix corresponding to
$\psi^G (\psi^H)$.
Then one can express the NIM-rep matrix $\hat{n}_{(\mu; \nu)}$
of the $G/H$ theory in terms of
$n^G_\mu$ and $n^H_\nu$ as follows
\beq
 (\hat{n}_{(\mu; \nu)})_{(\alpha; \beta)}{}^{(\alpha';\beta')} =
         \sum_{(J, J') \in \Gid(\V)}
         (n^G_\mu)_{J \alpha}{}^{\alpha'}
         (n^H_\nu)_{J' \beta}{}^{\beta'} .
\label{eq:finalNIM}
\eeq
For the regular NIM-reps, $\psi^G = S^G, \, \psi^H = S^H,$
one obtains $\Gid(\E) = \Gid(\V) = \Gid$ and
$\hat{\psi}$ coincides with $\hat{S}$.

In \cite{Ishikawa,IT}, the condition
$\tilde{b}^G_\alpha(J) = \tilde{b}^H_\beta(J')$
is called the brane selection rule, while
$(J \alpha; J' \beta) = (\alpha; \beta)$ is called
the brane identification.
If there is no fixed point in the brane identification,
the matrix \eqref{eq:finalpsi} yields a NIM-rep
$\hat{n}_{(\mu; \nu)}$ of the $G/H$ theory.
If we have fixed points in the brane identification,
we have to resolve them in order to obtain a unitary $\hat{\psi}$.
This is the same situation as the field identification
fixed points \cite{FSS2}.

\section{Mutual consistency of twisted boundary conditions}
\label{sec:compatibility}

As we have seen in the previous section,
the problem of solving the Cardy conditions is equivalent with finding
an appropriate NIM-rep of the fusion algebra.
However, the number of solutions may be greater than one
and we may have several NIM-reps for one particular fusion algebra.
Although these different NIM-reps are in general considered to be
corresponding to different modular invariants,
it is possible that two or more NIM-reps coexist in the
same modular invariant and are compatible with each other.
This issue of mutual consistency of different NIM-reps
has been studied in \cite{BFS,GG} (see also \cite{Ishikawa})
for the case of the boundary states twisted by an automorphism
of the chiral algebra.
We briefly review this in the following and discuss
its application to the case of coset models.

\bigskip

We begin with the relation of the twisted Ishibashi state
$\dket{\mu}_\omega$
for $\omega \neq 1$ with the untwisted one
$\dket{\mu}$.
Introducing an orthonormal basis $\{\ket{\varphi}\}$ of $\HH_\mu$,
the untwisted Ishibashi state can be written as
\beq
  \dket{\mu} = \sum_{\varphi}
  \ket{\varphi} \otimes \overline{\ket{\varphi}} \in
  \HH_\mu \otimes \widetilde{\HH}_{\bar{\mu}} .
\label{eq:Ishibashi}
\eeq
One can construct the Ishibashi states for $\omega \neq 1$
starting from $\dket{\mu}$.
Let $\Omega$ be a unitary operator mapping $\HH_\mu$
to $\HH_{\omega \mu}$,
\beq
  \Omega : \, \HH_\mu \rightarrow \HH_{\omega \mu} , \quad
  \Omega W_{n} \Omega^\dagger = \omega(W_{n}) , \quad
  \Omega \widetilde{W}_{n} \Omega^\dagger = \widetilde{W}_n .
\eeq
Then the twisted Ishibashi state $\dket{\mu}_\omega$
can be written as
\beq
  \dket{\mu}_\omega = \Omega \dket{\mu}_{\omega = 1}
  = \sum_{\varphi} (\Omega \ket{\varphi}) \otimes \overline{\ket{\varphi}} \in
  \HH_{\omega \mu} \otimes \widetilde{\HH}_{\overline{\mu}} .
\eeq
Actually, one can check that
this state satisfies the twisted gluing condition,
\bes
  (\omega(W_n) - (-1)^h \widetilde{W}_{-n}) \dket{\mu}_\omega
  &=
  (\Omega W_n \Omega^\dagger - (-1)^h \widetilde{W}_{-n})
   \Omega \dket{\mu}_{\omega = 1} \\
  &= \Omega (W_n - (-1)^h \widetilde{W}_{-n}) \dket{\mu}_{\omega = 1} \\
  &= 0 .
\ees
The normalization of $\dket{\mu}_\omega$ follows from that of $\dket{\mu}$
and coincides with our convention \eqref{eq:Ishibashinorm},
\beq
   {}_\omega\dbra{\mu} q^{H_c} \dket{\mu'}_\omega
   = \dbra{\mu} \Omega^\dagger q^{H_c} \Omega  \dket{\mu'}
   = \dbra{\mu} q^{H_c} \dket{\mu'}
   = \delta_{\mu\mu'} \chi_{\mu}(\tau) .
\eeq
We will also need the overlap of the twisted Ishibashi state with
the untwisted one.
Since $\dket{\mu}_\omega$ couples $\omega \mu$ with $\bar{\mu}$,
we have a non-vanishing overlap if and only if $\mu \in \I^\omega$,
\beq
  \dbra{\mu} q^{H_c} \dket{\mu'}_\omega
  = \dbra{\mu} q^{H_c} \Omega \dket{\mu'}
  = \begin{cases}
   \chi^\omega_\mu(\tau)
  &\text{if $\mu = \mu' \in \I^\omega$} , \\
   0
  &\text{otherwise} .
   \end{cases}
\label{eq:twin}
\eeq
Here $\chi^\omega_\mu$ is the twining character \cite{FSS1}
\beq
   \chi^\omega_\mu(\tau) = \trace_{\HH_\mu} \Omega q^{L_0 - \frac{c}{12}} .
\eeq
For the case of (untwisted) affine Lie algebras,
the twining character coincides with the ordinary character
of a twisted affine Lie algebra called the orbit Lie algebra \cite{FSS1}.

In order to obtain the spectrum in the open string channel,
we need the modular transformation of the twining character.
Since the twining character appears in the overlap between twisted
and untwisted states, its modular transformation is written in terms
of the characters of the twisted chiral algebra $\A^\omega$.
Here we denote by $\A^\omega$ an algebra generated by the currents
with the twisted boundary condition $W(e^{2\pi i}z) = \omega(W(z))$.
Let $\tilde{\I}$ be the set of all the possible representations of
the twisted chiral algebra $\A^\omega$
and $\chi_{\tilde{\mu}} \,(\tilde{\mu} \in \tilde{\I})$ be
the corresponding character.
Then the modular transformation of the twining character can be
written as
\begin{subequations}
\label{eq:St}
\bea
  \chi^\omega_\mu(-1/\tau) &= \sum_{\tilde{\mu} \in \tilde{\I}}
  \tilde{S}^{(0)}_{\mu \tilde{\mu}} \, \chi_{\tilde{\mu}}(\tau/r) \quad\quad
  (\mu \in \I^\omega), \\
  \chi_{\tilde{\mu}}(-1/\tau) &= \sum_{\mu \in \I^\omega}
  \tilde{S}^{(1)}_{\tilde{\mu} \mu} \, \chi^\omega_{\mu}(\tau/r) \quad\quad
  (\tilde{\mu} \in \tilde{\I}),
\eea
\end{subequations}
where $\tilde{S}^{(0)}$ and $\tilde{S}^{(1)}$ are
some unitary matrices \cite{BFS}.
We divide the argument of $\chi_{\tilde{\mu}}$ by the order $r$ of
the automorphism $\omega$ so that $\chi_{\tilde{\mu}}(\tau)$ is expanded
in integer powers of $q = e^{2\pi i \tau}$.
The properties of $\tilde{S}^{(0)}$ and $\tilde{S}^{(1)}$ have been
investigated in \cite{BFS}.
In particular, it has been shown that $\tilde{S}^{(1)}$
is related with $\tilde{S}^{(0)}$ via transposition,
\beq
 \tilde{S}^{(1)}_{\tilde{\mu} \mu} =
 \eta_\mu^{-2} \tilde{S}^{(0)}_{\mu \tilde{\mu}} ,
\label{eq:S0S1}
\eeq
where $\eta_\mu$ is an appropriate phase factor.

If $\A$ is an (untwisted) affine Lie algebra $\g^{(1)}$ at level $k$,
the corresponding twisted chiral algebra $\A^\omega$ is the twisted
affine Lie algebra $\g^{(r)}$ at level $k$ and
$\tilde{\I} = P_+^k(\g^{(r)})$.
\begin{table}
\begin{center}
\begin{tabular}{|c|ccccc|} \hline
$\g^{(r)}$            &  $A_{2l}^{(2)}$  &  $A_{2l-1}^{(2)}$
&  $D_{l+1}^{(2)}$   &  $E_{6}^{(2)}$   & $D_4^{(3)}$ \\ \hline
$\tilde{\g}^{(r)}$    &  $A_{2l}^{(2)}$  &  $D_{l+1}^{(2)}$
&  $A_{2l-1}^{(2)}$  &  $E_{6}^{(2)}$   & $D_{4}^{(3)}$   \\ \hline
\end{tabular}
\end{center}
\caption{The modular transformation of the twisted affine Lie algebras.
$l$ stands for the rank of the algebras.}
\label{tab:modular}
\end{table}
Since the modular transformation of the character of $\g^{(r)}$
is expressed by the characters of another affine Lie algebra
$\tilde{\g}^{(r)}$ \cite{Kac} (see Table~\ref{tab:modular}),
the matrices $\tilde{S}$ in \eqref{eq:St} are nothing but
the modular transformation matrix between $\g^{(r)}$ and
$\tilde{\g}^{(r)}$,
\footnote{%
This fact gives an explanation for the coincidence of the twining character
$\chi^\omega$ with the ordinary character of the orbit Lie algebra.}
for which
we can set $\eta_\mu = 1$ and obtain
$\tilde{S}^{(1)}{}^T = \tilde{S}^{(0)}$ \cite{Kac}.
In the following, we restrict ourselves to the case of $\eta_\mu = 1$
and write $\tilde{S}^{(0)}$ as $\tilde{S}$,
\beq
  \tilde{S}^{(0)} = \tilde{S}^{(1)}{}^T \equiv \tilde{S} .
\label{eq:assumeStilde}
\eeq
The matrix $\tilde{S}$ satisfies the following properties
\bes
  \tilde{S} \tilde{S}^\dagger &= 1,  \\
  (\tilde{S} \tilde{S}^T)_{\mu \nu} &= C_{\mu \nu} =  \delta_{\mu \bar{\nu}},
  \\
  (\tilde{S}^T \tilde{S})_{\tilde{\mu} \tilde{\nu}} &=
  \widetilde{C}_{\tilde{\mu} \tilde{\nu}} =
  \delta_{\tilde{\mu} \overline{\tilde{\nu}}} \quad\quad
  (\mu, \nu \in \I^\omega, \, \tilde{\mu}, \tilde{\nu} \in \tilde{\I}),
\ees
where $C$ and $\widetilde{C}$ are the charge-conjugation matrix
of $\I$ and $\tilde{\I}$, respectively.

\bigskip
We turn to the construction of the twisted Cardy states \cite{BFS,GG}.
Suppose that we obtain the twisted Cardy states
$\ket{\tilde{\alpha}}_\omega$ in the diagonal modular invariant,
\beq
  \ket{\tilde{\alpha}}_\omega = \sum_{\mu \in \I^\omega}
  \psi_{\tilde{\alpha}}{}^\mu \frac{1}{\sqrt{S_{0 \mu}}}
  \dket{\mu}_\omega .
\eeq
For $\ket{\tilde{\alpha}}_\omega$
to be well-defined in the diagonal modular invariant,
$\ket{\tilde{\alpha}}_\omega$
has to be consistent with the regular Cardy states
$\ket{\mu} \,(\mu \in \I)$.
From eqs.\eqref{eq:twin}\eqref{eq:St}, one can calculate
the overlap of the twisted states $\ket{\tilde{\alpha}}_\omega$
with the untwisted state $\ket{0} \,(0 \in \I)$,
\beq
  \bra{0} \tilde{q}^{H_c} \ket{\tilde{\alpha}}_\omega =
  \sum_{\mu \in \I^\omega} \psi_{\tilde{\alpha}}{}^\mu \,
  \chi^\omega_\mu(-1/\tau) =
  \sum_{\mu \in \I^\omega, \, \tilde{\mu} \in \tilde{\I}}
  \psi_{\tilde{\alpha}}{}^\mu \tilde{S}_{\mu \tilde{\mu}}
  \, \chi_{\tilde{\mu}}(\tau/r) .
\eeq
Clearly, this amplitude is made to be well-defined if one sets
$\psi = \tilde{S}^{-1}$ or $\tilde{S}^T$.
Here we take the latter choice $\psi = \tilde{S}^T$,
\beq
  \ket{\tilde{\mu}}_\omega = \sum_{\mu \in \I^\omega}
  \tilde{S}_{\tilde{\mu} \mu}  \frac{1}{\sqrt{S_{0 \mu}}}
  \dket{\mu}_\omega  \quad (\tilde{\mu} \in \tilde{\I}),
\label{eq:Cardy_t}
\eeq
for which the overlap with $\ket{0}$ reads
\beq
  \bra{0} \tilde{q}^{H_c} \ket{\tilde{\mu}}_\omega =
  \sum_{\tilde{\nu} \in \tilde{\I}}
  (\tilde{S}^T \tilde{S})_{\tilde{\mu}\tilde{\nu}}
  \chi_{\tilde{\nu}}(\tau/r) =
  \chi_{\overline{\tilde{\mu}}}(\tau/r) .
\label{eq:0t}
\eeq
The twisted Cardy states are therefore labeled by the representation
$\tilde{\mu} \in \tilde{\I}$ of the twisted chiral algebra $\A^\omega$.

The twisted state $\ket{\tilde{\mu}}_\omega$ is characterized
by the property that
the open string spectrum \eqref{eq:0t} between  $\ket{\tilde{\mu}}_\omega$
and $\ket{0}$ consists of $\chi_{\tilde{\mu}}$ only.
This is completely parallel with Cardy's original construction of
the untwisted Cardy states:
$\ket{\mu} \, (\mu \in \I)$ is determined so that
$\bra{\mu} \tilde{q}^{H_c} \ket{0} = \chi_\mu$ holds
\cite{Cardy}.
From this construction, Cardy has argued that the overlap
of the Cardy states can be expressed in terms of the fusion coefficients,
namely,
\beq
  \bra{\mu} \tilde{q}^{H_c} \ket{\bar{\nu}} =
  \sum_{\rho \in \I} \fusion{\mu}{\nu}{\rho} \chi_\rho(\tau) .
\label{eq:boundaryfusion}
\eeq
Substituting $\psi_{\mu}{}^{\nu} = S_{\mu \nu}$ in the left-hand side,
one obtains the Verlinde formula \eqref{eq:Verlinde}.
We can proceed in the same way for the twisted states \cite{GG}.
For simplicity, we restrict ourselves to the case of $r=2$.\footnote{%
For the case of $r>2$, we have to introduce $r-1$ twisted sectors in order
to describe the fusion rule of the twisted representations.}
Then the counterpart of \eqref{eq:boundaryfusion} can be written as
\begin{subequations}
\label{eq:boundaryfusion_t}
\bea
  {}_\omega\bra{\tilde{\mu}} \tilde{q}^{H_c} \ket{\bar{\nu}} &=
  \sum_{\tilde{\rho} \in \tilde{\I}} \fusion{\tilde{\mu}}{\nu}{\tilde{\rho}}
  \chi_{\tilde{\rho}}(\tau/r) ,\\
  {}_\omega\bra{\tilde{\mu}} \tilde{q}^{H_c} \ket{\bar{\tilde{\nu}}}_\omega
  &=
  \sum_{\rho \in \I} \fusion{\tilde{\mu}}{\tilde{\nu}}{\rho}
  \chi_{\rho}(\tau) ,
\eea
\end{subequations}
where $\fusion{\tilde{\mu}}{\nu}{\tilde{\rho}}$
can be considered as describing the fusion rule
of the twisted representation with the ordinary one.
Substituting the expression \eqref{eq:Cardy_t} for the twisted states,
we obtain a generalized Verlinde formula
\beq
  \fusion{\tilde{\mu}}{\tilde{\nu}}{\rho} =
  \fusion{\tilde{\mu}}{\bar{\rho}}{\bar{\tilde{\nu}}} =
  \sum_{\sigma \in \I^\omega}
  \frac{%
  \tilde{S}_{\tilde{\mu}\sigma} \tilde{S}_{\tilde{\nu}{\sigma}}
  \overline{S_{\rho \sigma}} }{S_{0 \sigma}} .
\label{eq:Verlinde_t}
\eeq
Since the fusion coefficients take values in non-negative integers,
the above expression \eqref{eq:boundaryfusion_t} for the overlaps
implies that the mutual consistency of the twisted states
\eqref{eq:Cardy_t} with the regular states.

\begin{figure}[tb]
\begin{center}
\includegraphics[width=13cm]{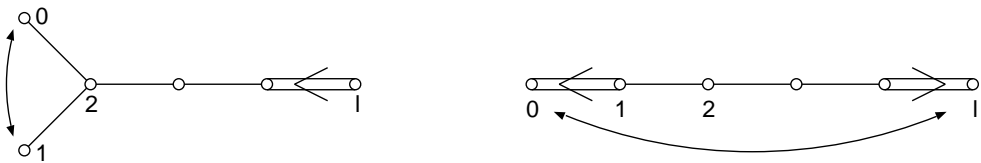}
\end{center}
\caption{The Dynkin diagrams of $A_{2l-1}^{(2)}$ (left) and
$D_{l+1}^{(2)}$ (right). The arrows stand for the action of
the diagram automorphisms.}
\label{fig:auto}
\end{figure}

The action of the simple currents on the twisted Cardy states
\eqref{eq:Cardy_t} can be described
in terms of the twisted chiral algebra $\A^\omega$.
In order to see this, consider the case of the affine Lie algebra $\g^{(1)}$
and the diagram automorphism $\omega$, for which
the coefficient $\tilde{S}_{\tilde{\mu} \nu}$ in the twisted Cardy states
is the modular transformation matrix between
the twisted affine Lie algebras $\g^{(r)}$ and $\tilde{\g}^{(r)}$
(see Table~\ref{tab:modular}).
The sets, $\V, \E$, of the labels of the Cardy states and the Ishibashi states,
respectively, read
\beq
 \V = \tilde{\I} = P_+^k(\g^{(r)}) , \quad
 \E = \I^\omega  \cong P_+^k(\tilde{\g}^{(r)}) .
\eeq
Here we used the equivalence of the twining character of $\g^{(1)}$
and the ordinary character of the orbit Lie algebra $\tilde{\g}^{(r)}$
to identify $\I^\omega$ with $P_+^k(\tilde{\g}^{(r)})$.
In this case, the action of the simple currents can be identified
with the diagram automorphisms of
$\g^{(r)}$ and $\tilde{\g}^{(r)}$ \cite{Ishikawa}, namely,
\beq
  \auto{\V} = \text{Aut}(\g^{(r)}) , \quad
  \auto{\E} \cong \text{Aut}(\tilde{\g}^{(r)}) ,
\eeq
where we denote by $\text{Aut}(\g^{(r)})$
the group of the diagram automorphisms of $\g^{(r)}$.
For $\g^{(r)} = A_{2l}^{(2)}, E_6^{(2)}$ and $D_4^{(3)}$,
there is no nontrivial automorphism and $\text{Aut}(\g^{(r)}) = \{1\}$.
For $\g^{(r)} = A_{2l-1}^{(2)}$ and $D_{l+1}^{(2)}$,
there is an automorphism of order $2$ and  $\text{Aut}(\g^{(r)}) \cong \Z_2$
(see Fig.~\ref{fig:auto}).

\bigskip
Let us apply the foregoing discussion to the case of the $G/H$ theory.
Suppose that
an automorphism $\omega$ of the horizontal  Lie algebra $\g$
leaves $\h \subset \g$ invariant.
Then $\omega$ induces an automorphism $\hat{\omega}$ of the $G/H$ theory,
which acts on $\hat{\I}$ as
\beq
  \hat{\omega} : (\mu; \nu) \mapsto (\omega \mu; \omega \nu) ,
  \quad (\mu; \nu) \in \hat{\I}.
\eeq
A natural candidate for the corresponding NIM-rep of $G/H$
can be constructed
starting from the twisted Cardy states \eqref{eq:Cardy_t} of
$G$ and $H$.
The formula \eqref{eq:finalpsi} yields the following diagonalization
matrix
\beq
  \hat{\psi}_{(\tilde{\mu}; \tilde{\nu})}{}^{(\mu; \nu)}
  = \sqrt{\abs{\Gid(\E)} \abs{\Gid(\V)}} \,
    \tilde{S}^{G}_{\tilde{\mu} \mu}
    \tilde{S}^{H}_{\tilde{\nu} {\bar{\nu}}} , \quad
  (\tilde{\mu}; \tilde{\nu}) \in \hat{\V} , \,
  (\mu; \nu) \in \hat{\E} ,
\label{eq:psihat_twisted}
\eeq
where $\hat{\V}$ and $\hat{\E}$ are defined as
\bea
  \hat{\V} &= \{(\tilde{\mu}; \tilde{\nu}) \, | \,
        \tilde{\mu} \in \tilde{\I}^G, \tilde{\nu} \in \tilde{\I}^H, \,
        \tilde{b}^G_{\tilde{\mu}}(J) = \tilde{b}^H_{\tilde{\nu}}(J') \,
        (\forall (J, J') \in \Gid(\E)), \notag \\
&\quad\quad\quad\quad\quad\quad\quad\quad\quad\quad\quad\quad
        (J \tilde{\mu}; J' \tilde{\nu}) = (\tilde{\mu}; \tilde{\nu}) \,
        (\forall (J, J') \in \Gid(\V)) \} ,
\label{eq:E_twisted}
\\
  \hat{\E} &= \{(\mu; \nu) \, | \,
        \mu \in \I^{G,\omega}, \nu \in \I^{H,\omega}, \,
        b^G_\mu(J) = b^H_\nu(J')  \,
        (\forall (J, J') \in \Gid(\V)), \notag \\
&\quad\quad\quad\quad\quad\quad\quad\quad\quad\quad\quad\quad
        (J \mu; J' \nu) = (\mu; \nu) \,
        (\forall (J, J') \in \Gid(\E)) \} .
\label{eq:V_twisted}
\eea
Since the spectrum $\hat{\E}$ of the Ishibashi states consists of
the elements invariant under $\hat{\omega}$,
one can expect that this NIM-rep describes the $\hat{\omega}$-twisted
Cardy states of $G/H$.
As we shall see below, this is the case if $\hat{\psi}$ in
\eqref{eq:psihat_twisted}
gives a well-defined NIM-rep.

We begin our discussion with the relation of the spectrum $\hat{\E}$
with the set
$\hat{\I}^{\hat{\omega}} \subset \hat{\I}$ of all the representations
fixed by $\hat{\omega}$,
\beq
  \hat{\I}^{\hat{\omega}} =
  \{(\mu; \nu) \in \hat{\I} \, | \, (\omega \mu; \omega \nu) = (\mu; \nu)\}.
\eeq
Clearly, $\hat{\E}$ is contained in $\hat{\I}^{\hat{\omega}}$;
the converse is in general not true.
Actually, it is possible that $(\omega \mu; \omega \nu)$ gives rise to
the same state as $(\mu; \nu)$ due to the field identification
even if $\omega \mu \neq \mu$ (or $\omega \nu \neq \nu$).
Therefore $\hat{\E}$ does not
necessarily coincide with $\hat{\I}^{\hat{\omega}}$;
rather a subset of $\hat{\I}^{\hat{\omega}}$ in general.
If $\hat{\E}$ is a true subset of $\hat{\I}^{\hat{\omega}}$,
we obtain several extra Ishibashi states.
This is a situation analogous to the case of orbifolds,
in which we may have some extra Ishibashi states from the twisted sectors.
These extra states
can be used to resolve the branes sitting at the fixed points
of the orbifold to yield fractional branes.
For the construction of the boundary states in the $G/H$ theory,
the counterpart of the orbifold fixed points
is the brane identification fixed point,
and one can expect that the existence of the extra Ishibashi states
implies the appearance of the brane identification fixed points,
which are to be resolved by the extra states.
As we will see later, this is checked for the $so(2n)$ diagonal coset.

In the rest of this section, we restrict ourselves to the case
of $\hat{\E} = \hat{\I}^{\hat{\omega}}$.
Then a $\hat{\omega}$-invariant
representation $(\mu; \nu) \in \hat{\I}^{\hat{\omega}}$ has a representative
such that $\mu \in \I^{G,\omega}, \nu \in \I^{H,\omega}$,
and the twining character of $(\mu; \nu)$ can be obtained
as the branching function of the twining character of $G$,
\begin{subequations}
\label{eq:cosettwining}
\bea
  \chi^{G,\omega}_\mu &=
  \sum_{\nu \in \I^{H,\omega}}
  \chi^{\hat{\omega}}_{(\mu;\nu)} \chi^{H,\omega}_\nu,
  \quad
  \mu \in \I^{G,\omega} . \\
\intertext{%
Similarly, the branching rule for the twisted chiral algebra
can be written as}
  \chi^{G}_{\tilde{\mu}} &=
  \sum_{\tilde{\nu} \in \tilde{\I}^H}
    \chi_{(\tilde{\mu};\tilde{\nu})} \chi^{H}_{\tilde{\nu}},
  \quad
  \tilde{\mu} \in \tilde{\I}^{G} .
\label{eq:cosettwining2}
\eea
\end{subequations}
For an affine Lie algebra $\g^{(1)}$ and
its diagram automorphism $\omega$ of order $r$,
the above equations can be regarded as the branching rule of
the orbit Lie algebra $\tilde{\g}^{(r)}$ and
the twisted Lie algebra $\g^{(r)}$, respectively.
The modular transformation of the twining character
$\chi^{\hat{\omega}}_{(\mu;\nu)}$ yields
$\chi_{(\tilde{\mu};\tilde{\nu})}$, and vice versa,
\begin{subequations}
\bea
  \chi^{\hat{\omega}}_{(\mu; \nu)}(-1/\tau) &=
  \sum_{(\tilde{\mu};\tilde{\nu}) \in \hat{\V}}
  \tilde{S}^{(0)}_{(\mu;\nu) (\tilde{\mu};\tilde{\nu})}
  \chi_{(\tilde{\mu};\tilde{\nu})}(\tau/r) , \\
  \chi_{(\tilde{\mu};\tilde{\nu})}(-1/\tau) &=
  \sum_{(\mu;\nu) \in \hat{\E}}
  \tilde{S}^{(1)}_{(\tilde{\mu};\tilde{\nu})(\mu;\nu)}
  \chi^{\hat{\omega}}_{(\mu; \nu)}(\tau/r) .
\eea
\end{subequations}
From the branching rules \eqref{eq:cosettwining},
one can express these modular transformation matrices
in terms of $\tilde{S}^G$ and $\tilde{S}^H$ in \eqref{eq:St},
\begin{subequations}
\label{eq:St_coset}
\bea
  \tilde{S}^{(0)}_{(\mu;\nu) (\tilde{\mu};\tilde{\nu})} &=
  \abs{\Gid(\V)} \tilde{S}^{G(0)}_{\mu \tilde{\mu}}
                 \overline{\tilde{S}^{H(0)}_{\nu \tilde{\nu}}} , \\
  \tilde{S}^{(1)}_{(\tilde{\mu};\tilde{\nu})(\mu;\nu)} &=
  \abs{\Gid(\E)} \tilde{S}^{G(1)}_{\tilde{\mu} \mu}
                 \overline{\tilde{S}^{H(1)}_{\tilde{\nu} \nu}} ,
\eea
\end{subequations}
where the factors, $\abs{\Gid(\V)}$, $\abs{\Gid(\E)}$, arise
from rewriting the sum over $\hat{\V}$ and $\hat{\E}$, respectively,
\begin{subequations}
\bea
  \sum_{(\tilde{\mu}; \tilde{\nu}) \in \hat{\V}} &\rightarrow \quad
     \frac{1}{\abs{\Gid(\V)}}
     \sum_{(\tilde{\mu}, \tilde{\nu}) \in \tilde{\I}^G \otimes \tilde{\I}^H}
     \frac{1}{\abs{\Gid(\E)}} \sum_{(J, J') \in \Gid(\E)}
            \tilde{b}^G_{\tilde{\mu}}(J) \tilde{b}^H_{\tilde{\nu}}(J')^{-1}, \\
  \sum_{(\mu; \nu) \in \hat{\E}} &\rightarrow \quad
     \frac{1}{\abs{\Gid(\E)}}
     \sum_{(\mu, \nu) \in \I^{G,\omega} \otimes \I^{H,\omega}}
     \frac{1}{\abs{\Gid(\V)}} \sum_{(J, J') \in \Gid(\V)}
            b^G_{\mu}(J) b^H_{\nu}(J')^{-1} .
\label{eq:sumEhat}
\eea
\end{subequations}
The expression \eqref{eq:St_coset} together with
the relation \eqref{eq:S0S1} between $\tilde{S}^{(0)}$ and $\tilde{S}^{(1)}$
implies that the order of the identification groups for $\V$ and $\E$
should take the same value $N$,
\beq
  \abs{\Gid(\V)} = \abs{\Gid(\E)} \equiv N .
\label{eq:N}
\eeq
Two modular transformation matrices in \eqref{eq:St_coset}
are therefore related with each other via transposition,
\beq
  \tilde{S}^{(0)}_{(\mu;\nu) (\tilde{\mu};\tilde{\nu})} =
  \tilde{S}^{(1)}_{(\tilde{\mu};\tilde{\nu}) (\mu;\nu)} =
  N \tilde{S}^{G}_{\mu \tilde{\mu}}
                 \overline{\tilde{S}^{H}_{\nu \tilde{\nu}}} ,
\eeq
where we used the relation \eqref{eq:assumeStilde}.

Clearly, the above matrix coincides with the diagonalization matrix
\eqref{eq:psihat_twisted} constructed from the twisted NIM-reps
of $G$ and $H$.
Hence the states $\ket{(\tilde{\mu};\tilde{\nu})}$
defined in \eqref{eq:psihat_twisted} are actually
the twisted Cardy states \eqref{eq:Cardy_t} for the automorphism
$\hat{\omega}$ of the $G/H$ theory, and
the set $\V$ of the labels for the Cardy states can be identified
with those for the branching functions \eqref{eq:cosettwining2}
of the twisted chiral algebra.

Applying the argument for the generic twisted states
to the present case, we can confirm
that the twisted states
$\ket{(\tilde{\mu};\tilde{\nu})} \,((\tilde{\mu};\tilde{\nu}) \in \hat{\V})$
are consistent with the regular states
$\ket{(\mu; \nu)} \,( (\mu; \nu) \in \hat{\I})$.
In particular, the overlaps can be expressed in exactly
the same manner as eq.\eqref{eq:Verlinde_t},
\begin{subequations}
\bea
  \bra{(\tilde{\mu};\tilde{\nu})}
     \tilde{q}^{H_c} \ket{(\bar{\mu};\bar{\nu})} &=
  \sum_{(\tilde{\mu}';\tilde{\nu}') \in \hat{\V}}
  \hat{\mathcal{N}}_{(\tilde{\mu};\tilde{\nu}) (\mu;\nu)}
      {}^{(\tilde{\mu}';\tilde{\nu}')}
  \chi_{(\tilde{\mu}';\tilde{\nu}')}(\tau/r) ,\\
  \bra{(\tilde{\mu};\tilde{\nu})}
     \tilde{q}^{H_c}
  \ket{(\overline{\tilde{\mu}'};\overline{\tilde{\nu}'})}
  &=
  \sum_{(\mu;\nu) \in \hat{\I}}
  \hat{\mathcal{N}}_{(\tilde{\mu};\tilde{\nu})(\tilde{\mu}';\tilde{\nu}')}
      {}^{(\mu;\nu)}
  \chi_{(\mu;\nu)}(\tau) .
\eea
\end{subequations}
The fusion coefficients for the twisted representations
take the form
\bes
 \hat{\mathcal{N}}_{(\mu; \nu) (\tilde{\mu}; \tilde{\nu})}
      {}^{(\tilde{\mu}';\tilde{\nu}')} &=
 \sum_{(\rho;\sigma) \in \hat{\E}}
     \frac{%
        \hat{S}_{(\mu;\nu)(\rho;\sigma)}
        \tilde{S}_{(\tilde{\mu}; \tilde{\nu})(\rho;\sigma)}
        \overline{\tilde{S}_{(\tilde{\mu}'; \tilde{\nu}')(\rho;\sigma)}}}{%
        \hat{S}_{(0;0)(\rho;\sigma)}} \\
  &= \sum_{(\rho, \sigma) \in \I^{G,\omega} \otimes \I^{H,\omega}}\,
     \sum_{(J, J') \in \Gid(\V)}
            b^G_{\rho}(J) b^H_{\sigma}(J')^{-1}
     \frac{%
        S^G_{\mu \rho}
        \tilde{S}^G_{\tilde{\mu}\rho}
        \overline{\tilde{S}^G_{\tilde{\mu}' \rho}}}{%
        S^G_{0 \rho}}
     \frac{%
        \overline{S^H_{\nu \sigma}}
        \overline{\tilde{S}^H_{\tilde{\nu} \sigma}}
        \tilde{S}^H_{\tilde{\nu}' \sigma}}{%
        \overline{S^H_{0 \sigma}}} \\
  &=       \sum_{(J, J') \in \Gid(\V)}
         \mathcal{N}^G_{\mu J\!\tilde{\mu}}{}^{\tilde{\mu}'}
         \mathcal{N}^H_{\nu J'\!\tilde{\nu}}{}^{\tilde{\nu}'} ,
\ees
where we used the formula \eqref{eq:sumEhat}
and the relation \eqref{eq:N}.
One can see that this expression for the overlaps
of the twisted Cardy states coincides with
the formula \eqref{eq:finalNIM} for generic NIM-reps of the $G/H$ theory.

\section{Boundary states in $so(2n)$ diagonal coset theory}
\label{sec:diagonal}

The diagonal coset theory $so(2n)_1 \oplus so(2n)_1/so(2n)_2$
has the central charge $c=1$ and is equivalent
to the orbifold $S^1/\Z_2$ at the radius $R=\sqrt{2n}$.
In this section, we apply the method given in the previous sections
to this particular model to construct the twisted Cardy states
following from the diagram automorphisms of $so(2n)$.
We will compare the result with the boundary states of $S^1/\Z_2$
in the next section.

\subsection{Preliminaries}

We denote by $\I_k$ the set of all the integrable representations of
$so(2n)_k$.
For $k=1,2$, we obtain
\begin{subequations}
\bea
  \I_1 &= \{\Lambda_0, \Lambda_1, \Lambda_{n-1}, \Lambda_n \} , \\
  \I_2 &= \{2\Lambda_0, 2\Lambda_1, 2\Lambda_{n-1}, 2\Lambda_n, \notag\\
  & \quad\quad
        \Lambda_0 + \Lambda_{n-1},  \Lambda_0 + \Lambda_{n},
        \Lambda_1 + \Lambda_{n-1},  \Lambda_1 + \Lambda_{n},
        \lambda_j \,(j=1,2,\dots,n-1) \},
\eea
\end{subequations}
where
$\Lambda_i$ is the $i$-th fundamental weight of the affine $so(2n)$
and $\lambda_j$ are defined as
\begin{equation}
\lambda_j=
 \begin{cases}
  \Lambda_0+\Lambda_1 & (j=1) ,\\
  \Lambda_j&(2\leq j \leq n-2) ,\\
  \Lambda_{n-1}+\Lambda_{n} & (j=n-1) .
 \end{cases}
\end{equation}
The simple current group of $so(2n)_k$ consists of four elements
\beq
  \G = \{J_O = k \Lambda_0 ,  J_V = k \Lambda_1,
         J_C = k \Lambda_{n-1}, J_S = k \Lambda_{n} \} .
\eeq
For even $n$, $\G$ is factorized as
\beq
  \G = \{J_O, J_C\} \times \{J_O, J_S \} \cong \Z_2 \times \Z_2
  \quad (n : \text{even}),
\eeq
while, for odd $n$, $\G$ is a cyclic group $\Z_4$
\beq
  \G = \{J_O, J_S, J_V = (J_S)^2, J_C = (J_S)^3 \} \cong \Z_4
  \quad (n : \text{odd}) .
\eeq
The modular transformation matrices together with the action of the
simple currents are found in Appendix~\ref{sec:S1}.

One can construct the primary fields of $so(2n)_1 \oplus so(2n)_1/so(2n)_2$
by applying the selection rule and the field identification
to $\I_1 \otimes \I_1 \otimes \I_2$.
Since we are considering the diagonal embedding
$so(2n)_2 \subset so(2n)_1 \oplus so(2n)_1$,
the identification currents are of the form $(J, J, J) \,(J \in \G)$
and $\Gid$ is isomorphic to $\G$.
There is no fixed points in the field identification and we have
$\abs{\hat{\I}} = 4 \times 4 \times (n+7)/
(4 \times 4) = n + 7$ primary fields.
We list the resulting spectrum in Table~\ref{tab:field},
in which each field is denoted by a single symbol.
\footnote{%
Our notation is the same as that of \cite{Cappelli}.
}
\begin{table}
\begin{center}
{\renewcommand{\arraystretch}{1.2}
\begin{tabular}[t]{|c|c|c|}
\hline
field & coset representation & weight\\
\hline
$u_+$   & $(\Lambda_0,\Lambda_0;2\Lambda_0)$ & $0$ \\
$u_-$   & $(\Lambda_0,\Lambda_0;2\Lambda_1)$ & $1$ \\
$\begin{gathered}
  \phi_+ \\ \phi_-
 \end{gathered}$ &
$ \begin{gathered}
        (\Lambda_0,\Lambda_0;2\Lambda_{n-1})\\
        (\Lambda_0,\Lambda_0;2\Lambda_{n})
  \end{gathered}
  \quad n : \text{even}\quad\quad
  \begin{gathered}
        (\Lambda_0,\Lambda_1;2\Lambda_{n-1})\\
        (\Lambda_0,\Lambda_1;2\Lambda_{n})
 \end{gathered}
  \quad n : \text{odd}$ & $n/4$  \\
$\chi_r$ &
$ \begin{cases}
  (\Lambda_0,\Lambda_0;\lambda_{r})& r : \text{even}\\
  (\Lambda_0,\Lambda_1;\lambda_{r})& r : \text{odd}
\end{cases}$ &
$r^2/4n$ \\
$\sigma_0$ & $(\Lambda_0,\Lambda_{n-1};\Lambda_{0}+\Lambda_{n-1})$ & $1/16$\\
$\sigma_1$ & $(\Lambda_0,\Lambda_{n};\Lambda_{0}+\Lambda_{n})$ & $1/16$\\
$\tau_0$ & $(\Lambda_0,\Lambda_{n-1};\Lambda_{1}+\Lambda_{n})$ & $9/16$\\
$\tau_1$ & $(\Lambda_0,\Lambda_{n};\Lambda_{1}+\Lambda_{n-1})$ & $9/16$\\
\hline
\end{tabular}
}
\end{center}
\caption{The spectrum of the diagonal coset
$so(2n)_1 \oplus so(2n)_1/so(2n)_2$.
Here $r = 1,2,\dots,n-1$, and there are $n+7$ fields.
The last column shows the conformal
weight of each field.
}
\label{tab:field}
\end{table}
The characters of the coset theory can be obtained as
the branching function for the embedding
$so(2n)_2 \subset so(2n)_1 \oplus so(2n)_1$.
We list the explicit form of the characters in Appendix~\ref{sec:branching}.
Since there is no identification fixed points,
the modular transformation matrix $\hat{S}$
of the coset theory can be obtained from the formula \eqref{eq:cosetSmatrix},
\bea
  \hat{S} = &\frac{1}{\sqrt{8n}}\times \notag\\
    & {\renewcommand{\arraystretch}{1.2}
     \begin{array}{c|ccccc}
          & u_\pm & \phi_\pm & \chi_s & \sigma_k & \tau_k \\
      \hline
    u_\pm & 1 &  1 & 2 & \pm\sqrt{n} & \pm\sqrt{n}  \\
    \phi_\pm & 1 & (-1)^n & 2(-1)^s &
              \pm (-i)^n (-1)^k \sqrt{n} & \pm (-i)^n (-1)^k \sqrt{n} \\
    \chi_r & 2 & 2(-1)^r & 4\cos \frac{\pi r s}{n} & 0 & 0 \\
    \sigma_j& \pm\sqrt{n}& \pm (-i)^n (-1)^j \sqrt{n} & 0 &
              (M^n)_{jk}\sqrt{2n}& -(M^n)_{jk}\sqrt{2n} \\
    \tau_j  & \pm\sqrt{n}& \pm (-i)^n (-1)^j \sqrt{n} & 0 &
             -(M^n)_{jk}\sqrt{2n}&  (M^n)_{jk}\sqrt{2n}
     \end{array} }  \notag \\
    & \quad \text{where} \quad
         M = \frac{1}{\sqrt{2}}
         \begin{pmatrix}
           e^{- i \pi/4}& e^{ i \pi/4}\\
           e^{ i \pi/4} & e^{ -i \pi/4}
         \end{pmatrix} , \quad M^4 = 1.
\label{eq:Shat}
\eea

\subsection{Construction of the twisted Cardy states}

As we have seen in the previous sections,
we can construct the Cardy states, or NIM-reps, of the $G/H$ theory starting
from those of the $G$ and $H$ theories.
By taking non-trivial NIM-reps of $G$ or $H$, one could in general
obtain a non-trivial NIM-rep of $G/H$, besides the regular one.
In particular, the Cardy states twisted by an automorphism
of the $G/H$ chiral algebra follow from those of the $G$ and $H$ theories.
As is argued in Section~\ref{sec:compatibility}, these twisted states
are consistent with the regular ones
and can be considered as the states in the diagonal modular invariant.

For the $so(2n)$ diagonal coset theory, the relevant chiral algebras are
$\g = so(2n)_1 \oplus so(2n)_1$ and $\h = so(2n)_2$.
There are three types of automorphisms available
for twisting the boundary condition of the coset theory:
the order-$2$ automorphism $\omega_2$ of $so(2n)$,
the order-$3$ automorphism $\omega_3$ of $so(8)$ and
the permutation automorphism $\pi$ of $so(2n) \oplus so(2n)$.
As we will see below, we can obtain twisted states for $\omega_2$
and $\omega_3$ by applying the construction of the previous sections.
For the permutation automorphism $\pi$, however,
the resulting NIM-rep suffers from the brane identification fixed points,
and we obtain the known NIM-reps after the resolution of the fixed points.

\subsubsection{$\Z_2$-automorphism of $so(2n)$}

The order-$2$ automorphism $\omega_2$ of $D_n = so(2n)$ exchanges
two short legs of the Dynkin diagram of $D_n$,
\beq
  \omega_2 : \Lambda_{n-1} \leftrightarrow \Lambda_n , \quad
             \Lambda_i \mapsto \Lambda_i \, (i \neq n-1, n) .
\eeq
From this automorphism, one obtains an automorphism
$\omega_2 \oplus \omega_2$ of $\g = so(2n)_1 \oplus so(2n)_1$,
which clearly makes invariant $\h = so(2n)_2 \subset \g$
and yields an automorphism $\hat{\omega}_2$ of the coset theory,
\beq
  \hat{\omega}_2 : (\mu_1, \mu_2; \nu) \mapsto
                   (\omega_2 \mu_1, \omega_2 \mu_2; \omega_2 \nu) , \quad
  (\mu_1, \mu_2; \nu) \in \hat{\I}.
\eeq
Using the notation in Table~\ref{tab:field}, the action of $\hat{\omega}_2$
can be written as
\beq
  \hat{\omega}_2 : \{u_\pm, \phi_\pm, \chi_r, \sigma_j, \tau_j\} \mapsto
                   \{u_\pm, \phi_\mp, \chi_r, \sigma_{1-j}, \tau_{1-j}\} .
\label{eq:omega2}
\eeq
The building blocks of the construction are the twisted NIM-rep
of $so(2n)_1$ and $so(2n)_2$.
As is explained in Section~\ref{sec:compatibility}
(in particular eq.\eqref{eq:Cardy_t}),
the diagonalization matrix of the twisted NIM-rep is given by
the modular transformation matrix of the twisted chiral algebra,
which is $D_n^{(2)}$ in the present case.
We list the necessary facts about $D_n^{(2)}$ in Appendix~\ref{sec:S2}.

The NIM-rep for $\hat{\omega}_2$ is obtained from
the formula \eqref{eq:psihat_twisted} by taking
$\tilde{S}^G = \tilde{S}_1 \otimes \tilde{S}_1$ and
$\tilde{S}^H = \tilde{S}_2$, where $\tilde{S}_k$ stands for
the modular transformation matrix of $D_n^{(2)}$ at level $k$.
The spectrum $\hat{\E}$ of the Ishibashi states follows from
the formula \eqref{eq:E_twisted},
\beq
  \hat{\E} = \{(\Lambda_0, \mu; \nu) \in \hat{\I} \, | \,
        \mu \in \I^{\omega_2}_1, \nu \in \I^{\omega_2}_2 \}
  = \{u_\pm, \chi_r \,(r = 1,2,\dots, n-1)\},
\eeq
where we denote by $\I^{\omega_2}_k$ the set of the $\omega_2$-invariant
representations of $so(2n)_k$.
One can easily confirm from the map \eqref{eq:omega2}
that $\hat{\E}$ coincides with the set $\hat{\I}^{\hat{\omega}_2}$
of all the $\hat{\omega}_2$-invariant representations and that
there is no extra Ishibashi states.

The Cardy states
are labeled by the representations of $D_n^{(2)}$.
The set $\hat{\V}$ of the Cardy states
follows from the formula \eqref{eq:V_twisted},
\bes
  \hat{\V} &= \{(\tilde{\mu}_1, \tilde{\mu}_2; \tilde{\nu}) \, | \,
        \tilde{\mu}_1, \tilde{\mu}_2 \in \tilde{\I}_1;
        \tilde{\nu} \in \tilde{\I}_2; \,
        \tilde{b}_{\tilde{\mu}_1} \tilde{b}_{\tilde{\mu}_2}
        = \tilde{b}_{\tilde{\nu}};  \, \\
&\quad\quad\quad\quad\quad\quad\quad\quad
        (J \tilde{\mu}_1, J \tilde{\mu}_2; J \tilde{\nu})
         = (\tilde{\mu}_1,\tilde{\mu}_2; \tilde{\nu}) \,
        (\forall (J, J, J) \in \Gid(\V)) \} ,
\label{eq:Dn2V_temp}
\ees
where we denote by $\tilde{I}_k$ the set of the integrable representations
of $D_n^{(2)}$ at level $k$.
The identification current groups, $\Gid(\E)$ and $\Gid(\V)$, can
be regarded as the group of the diagram automorphisms
of $A_{2n-3}^{(2)}$ and $D_n^{(2)}$, respectively.
$\tilde{b}_{\tilde{\mu}}$ in \eqref{eq:Dn2V_temp} is the monodromy charge for
the action of $\Gid(\E)$ (see Appendix~\ref{sec:S2}),
whereas $J$ stands for the diagram automorphism of $D_n^{(2)}$.
Since there is no fixed points in $\tilde{\I}_1$, we have also
no fixed points in the brane identification.
Consequently, one can identify $\hat{\V}$ with the set of
the labels for branching functions of
$D_{n, 2}^{(2)} \subset D_{n, 1}^{(2)} \oplus D_{n, 1}^{(2)}$,
which consists of the following elements
\beq
  \hat{\V} = \{ (\tilde{\Lambda}_0, \tilde{\Lambda}_0; \tilde{\lambda}_r) \,
   (r = 0,1,\dots, n-1),
   (\tilde{\Lambda}_0, \tilde{\Lambda}_{n-1};
    \tilde{\Lambda}_0 + \tilde{\Lambda}_{n-1}) \} .
\eeq
Here $\tilde{\Lambda}_j$ are the fundamental weights of $D_n^{(2)}$
and $\tilde{\lambda}_j \in \tilde{\I}_2$ are defined in Appendix~\ref{sec:S2}.
Putting these things together,
the diagonalization matrix \eqref{eq:psihat_twisted} can be written
as follows,
\beq
  \psi^{\hat{\omega}_2}  =
     \frac{1}{\sqrt{2n}}\times {\renewcommand{\arraystretch}{1.5}
           \begin{array}{c|cc}
                 &u_\pm & \chi_s \\ \hline
   (\tilde{\Lambda}_0,\tilde{\Lambda}_0;\tilde{\lambda}_r)
           &1 &2\cos(\frac{\pi}{2n} (2r+1)s ) \\
   (\tilde{\Lambda}_0,\tilde{\Lambda}_{n-1};
    \tilde{\Lambda}_0 + \tilde{\Lambda}_{n-1}) & \pm\sqrt{n} & 0
           \end{array} }
\label{eq:psihat_omega2}
\eeq
We have $n+1$ twisted Cardy states for $\hat{\omega}_2$.
The overlaps of these states with the regular ones
are expanded into the branching functions of $D_n^{(2)}$ (See
Appendix~\ref{sec:bt2} for the explicit form of the branching functions.)
For example, the overlap of the twisted state
$\ket{(\tilde{\Lambda}_0,\tilde{\Lambda}_0;\tilde{\lambda}_r)}$
with the regular state $\ket{u_+}$ can be written as
\beq
  \bra{(\tilde{\Lambda}_0,\tilde{\Lambda}_0;\tilde{\lambda}_r)}
  \tilde{q}^{H_c} \ket{u_+} =
  \chi_{(\tilde{\Lambda}_0,\tilde{\Lambda}_0;\tilde{\lambda}_r)}(\tau/2),
\eeq
where
$\chi_{(\tilde{\Lambda}_0,\tilde{\Lambda}_0;\tilde{\lambda}_r)}$ is
the branching function of
$D_{n, 2}^{(2)} \subset D_{n, 1}^{(2)} \oplus D_{n, 1}^{(2)}$.

\subsubsection{$\Z_3$-automorphism of $so(8)$}

The case of the order-3 automorphism $\omega_3$ of $D_4 = so(8)$
can be treated in the same way as $\omega_2$.
The automorphism $\omega_3$ of $D_4$ permutes
three legs of the Dynkin diagram of $D_4$,
\beq
  \omega_3 : \Lambda_1 \rightarrow \Lambda_3 \rightarrow \Lambda_4
                       \rightarrow \Lambda_1 .
\eeq
This induces an automorphism of the coset theory, which we denote by
$\hat{\omega}_3$,
\beq
  \hat{\omega}_3 : (\mu_1, \mu_2; \nu) \mapsto
                   (\omega_3 \mu_1, \omega_3 \mu_2; \omega_3 \nu) , \quad
  (\mu_1, \mu_2; \nu) \in \hat{\I}.
\eeq
There are $4 + 7 = 11$ representations for the $so(8)$ coset theory.
Using the notation in Table~\ref{tab:field}, the action of $\hat{\omega}_3$
can be written as
\beq
  \hat{\omega}_3 :
  \left\{
  \begin{matrix}
     u_+; & \chi_2; \\
     u_-, & \phi_+, & \phi_-; \\
     \chi_1,  & \sigma_0, & \sigma_1; \\
     \chi_3,  & \tau_0,   & \tau_1
  \end{matrix}\right\}
     \mapsto
  \left\{
  \begin{matrix}
     u_+; & \chi_2; \\
     \phi_+, & \phi_-, & u_-; \\
     \sigma_0, & \sigma_1, & \chi_1; \\
     \tau_0,   & \tau_1 ,  & \chi_3
  \end{matrix}\right\} .
\label{eq:omega3}
\eeq
One can see that there are three orbits of length $3$ together with
two fixed points $u_+, \chi_2$.

The twisted NIM-rep for $\omega_3$ is given by
the modular transformation matrix of the twisted chiral algebra $D_4^{(3)}$.
We list the necessary facts about $D_4^{(3)}$ in Appendix~\ref{sec:S3}.
We can obtain the NIM-rep for $\hat{\omega}_3$ of the coset theory
applying the formula \eqref{eq:psihat_twisted} to
$\tilde{S}^G = \tilde{S}_1 \otimes \tilde{S}_1$ and
$\tilde{S}^H = \tilde{S}_2$.
Here $\tilde{S}_k$ stands for
the modular transformation matrix of $D_4^{(3)}$ at level $k$.
The spectrum $\hat{\E}$ of the Ishibashi states follows from
the formula \eqref{eq:E_twisted},
\beq
  \hat{\E} = \{(\Lambda_0, \mu; \nu) \in \hat{\I} \, | \,
        \mu \in \I^{\omega_3}_1, \nu \in \I^{\omega_3}_2 \}
  = \{u_+, \chi_2 \} ,
\label{eq:Ehat_omega3}
\eeq
where we denote by $\I^{\omega_3}_k$ the set of the $\omega_3$-invariant
representations of $so(8)_k$.
One can see that
$\hat{\E}$ coincides with the set $\hat{\I}^{\hat{\omega}_3}$
of all the $\hat{\omega}_3$-invariant representations.

The Cardy states
are labeled by the representations of $D_4^{(3)}$.
Since there is no non-trivial diagram automorphism for $D_4^{(3)}$,
we have no non-trivial identification currents in the present case,
\textit{i.e.},
$\Gid(\V) = \Gid(\E) = \{1\}$.
Then the formula \eqref{eq:V_twisted} yields the set $\V$
in the form
\bes
  \hat{\V} &= \{(\tilde{\mu}_1, \tilde{\mu}_2; \tilde{\nu}) \, | \,
        \tilde{\mu}_1, \tilde{\mu}_2 \in \tilde{\I}_1;
        \tilde{\nu} \in \tilde{\I}_2 \} =
  \{(\tilde{\Lambda}_0, \tilde{\Lambda}_0; 2 \tilde{\Lambda}_0),
    (\tilde{\Lambda}_0, \tilde{\Lambda}_0; \tilde{\Lambda}_1) \},
\label{eq:D4V_temp}
\ees
where we denote by $\tilde{I}_k$ the set of the integrable representations
of $D_4^{(3)}$ at level $k$ and
by $\tilde{\Lambda}_j \,(j = 0,1,2)$ the fundamental weights of $D_4^{(3)}$.
$\hat{\V}$ can be regarded as the set of
the labels for the branching functions of
$D_{4, 2}^{(3)} \subset D_{4, 1}^{(3)} \oplus D_{4, 1}^{(3)}$.
The diagonalization matrix \eqref{eq:psihat_twisted} can be written
as follows,
\beq
  \psi^{\hat{\omega}_3} =
     \frac{1}{\sqrt{2}}\times {\renewcommand{\arraystretch}{1.5}
           \begin{array}{c|cc}
                 &u_+ & \chi_2 \\ \hline
   (\tilde{\Lambda}_0,\tilde{\Lambda}_0;2\tilde{\Lambda}_0)
           & 1 &  1 \\
   (\tilde{\Lambda}_0,\tilde{\Lambda}_0; \tilde{\Lambda}_1)
           & 1 & -1
           \end{array} }
\label{eq:psihat_omega3}
\eeq
We have two twisted Cardy states for $\hat{\omega}_3$.
The overlaps of these states with the regular ones
are expanded into the branching functions of $D_4^{(3)}$
in the manner parallel to the case of $\hat{\omega}_2$.

\subsubsection{Permutation automorphism of $so(2n) \oplus so(2n)$}

The permutation automorphism $\pi$ acts on
$\g = so(2n)_1 \oplus so(2n)_1$ as the exchange of two $so(2n)$ factors.
Clearly, $\pi$ leaves invariant $\h = so(2n)_2 \subset \g$
and induces an automorphism $\hat{\pi}$ of the coset theory,
\beq
  \hat{\pi} : (\mu_1, \mu_2; \nu) \mapsto (\mu_2, \mu_1; \nu) ,
  \quad (\mu_1, \mu_2; \nu) \in \hat{\I}.
\label{eq:pihat}
\eeq
In order to obtain the corresponding NIM-rep of the coset theory,
we need to know the NIM-rep of $so(2n)_1 \oplus so(2n)_1$
for the boundary condition twisted by the automorphism $\pi$.
This problem of finding a NIM-rep for the permutation automorphism
of tensor product theories has been solved in \cite{Recknagel},
which we briefly review.

Suppose that we have a NIM-rep
$(n_\mu)_{\alpha}{}^\beta \, (\mu \in \I; \alpha, \beta \in \V)$
of a fusion algebra $\I$ with the fusion coefficients
$\fusion{\mu}{\nu}{\rho}$.
Let $\psi_\alpha{}^\rho \,(\rho \in \E)$
be the corresponding diagonalization matrix,
$n_\mu = \psi \gamma^{(\mu)} \psi^\dagger$.
Then the problem is to find a NIM-rep $n^\pi$
for the tensor product algebra $\I \otimes \I$ with the spectrum
$\E^{\pi} = \{ (\mu, \mu) \, | \, \mu \in \E \}$,
which consists of the elements fixed by the permutation
$\pi : (\mu, \nu) \mapsto (\nu, \mu)$.
This is solved by setting
\beq
  n^\pi_{(\mu, \nu)} = n_\mu n_\nu ,
\label{eq:npi}
\eeq
where the product in the right-hand side is taken as
a $\abs{\V} \times \abs{\V}$ matrix.
Since $n_\mu$ form a NIM-rep of $\I$,
$n^\pi$ also has entries with non-negative integers.
Hence what we have to do is to check that $n^\pi$ satisfy
the fusion algebra $\I \otimes \I$,
which can be shown as follows,
\bes
  n^\pi_{(\mu, \nu)} n^\pi_{(\mu', \nu')}
   = (n_\mu n_\nu) (n_{\mu'} n_{\nu'})
  &= (n_\mu n_{\mu'}) (n_{\nu} n_{\nu'}) \\
  &= \sum_{\mu'' \in \I} \fusion{\mu}{\mu'}{\mu''} n_{\mu''}
    \sum_{\nu'' \in \I} \fusion{\nu}{\nu'}{\nu''} n_{\nu''} \\
  &= \sum_{(\mu'', \nu'') \in \I \otimes \I}
    \fusion{\mu}{\mu'}{\mu''} \fusion{\nu}{\nu'}{\nu''} \,
    n^\pi_{(\mu'', \nu'')} .
\ees
Here we used the fact that the fusion algebra is commutative
to change the order of the product.
The diagonalization matrix $\psi^\pi$ for $n^\pi$ is given by
$\psi$, which follows immediately
from the definition \eqref{eq:npi} of $n^\pi$,
\bes
  (n^\pi_{(\mu, \nu)})_{\alpha}{}^\beta
  = (n_\mu n_\nu)_\alpha{}^\beta
  &= (\psi \gamma^{(\mu)} \psi^\dagger
      \psi \gamma^{(\nu)} \psi^\dagger)_{\alpha}{}^\beta \\
  &= \sum_{\rho \in \E} \psi_\alpha{}^\rho
          \gamma^{(\mu)}_\rho \gamma^{(\nu)}_\rho
          \overline{\psi_\beta{}^\rho} \\
  &= \sum_{\rho \in \E} \psi_\alpha{}^\rho
          \gamma^{(\mu, \nu)}_{(\rho, \rho)}
          \overline{\psi_\beta{}^\rho} .
\ees
This calculation also shows that the spectrum for $n^\pi$ consists
of only the elements $(\rho, \rho) \in \E \otimes \E$
fixed by the permutation $\pi$.
Hence $n^\pi$ defined in \eqref{eq:npi} is actually a NIM-rep
of $\I \otimes \I$ with the spectrum $\E^\pi$.\footnote{%
In \cite{Recknagel}, only the case of the regular NIM-rep $n= N$ for $\I$
is considered. Our argument provides its generalization to
a generic NIM-rep $n \neq N$.
}
Note that the Cardy states for $n^\pi$ are also labeled by $\V$ for $n$.

The action of the simple currents \eqref{eq:Gpsi}
on the permutation NIM-rep $n^\pi$ follows from that on $n$.
First, the simple current group $\auto{\E^\pi}$
for the Ishibashi states is the diagonal subgroup
of $\auto{\E} \times \auto{\E}$,
\beq
  J : (\mu, \mu) \mapsto (J\mu, J\mu) , \quad J \in \auto{\E} ,
\eeq
which acts on $\psi^\pi$ as
\beq
  \psi^{\pi}_{\alpha}{}^{(J\mu, J\mu)}
  = \psi_{\alpha}{}^{J\mu}
  = \tilde{b}_{\alpha}(J) \psi_{\alpha}{}^{\mu}
  = \tilde{b}_{\alpha}(J) \psi^{\pi}_{\alpha}{}^{(\mu, \mu)} .
\eeq
The action on the Cardy states is given by
$(J, J') \in \auto{\V} \times \auto{\V}$,
\beq
  \psi^{\pi}_{\alpha}{}^{(\mu,\mu)} b_{(\mu, \mu)}((J,J'))
  = \psi^{\pi}_{\alpha}{}^{(\mu,\mu)} b_{\mu}(J) b_{\mu}(J')
  = \psi_{\alpha}{}^{\mu} b_{\mu}(J J')
  = \psi_{J J' \alpha}{}^{\mu}
  = \psi^{\pi}_{J J' \alpha}{}^{(\mu,\mu)},
\label{eq:sconVpi}
\eeq
where we used the fact that $J \mapsto b_\mu(J)$ is a representation
of the simple current group.
This action has a non-trivial stabilizer,
\bes
  \stab{\V^\pi} &= \{(J,J') \in \auto{\V} \times \auto{\V} \,|\,
  b_\mu(J J') = 1 \,\,\forall \mu \in \E \} \\
  &= \{(J,J') \in \auto{\V} \times \auto{\V} \,|\,
  J J' = 1 \} \\
  &\cong \auto{\V} ,
\ees
and the simple current group $\auto{\V^\pi}$ for the Cardy states
is defined as
\beq
  \auto{\V^\pi} = \auto{\V} \times \auto{\V} / \stab{\V^\pi}
  \cong \auto{\V} .
\eeq
The above construction of the permutation NIM-rep
is easily generalized to the case of
tensor products with more than two factors.
Namely, a NIM-rep corresponding to the permutation
automorphism $\pi$ is obtained as
\beq
  n^\pi_{\mu_1 \otimes \mu_2 \otimes \cdots \otimes \mu_n} =
  n_{\mu_1} n_{\mu_2} \cdots n_{\mu_n} .
\label{eq:npi2}
\eeq
The diagonalization matrix is again given by that for $n$.

Let us apply the above construction
to the $so(2n)_1 \oplus so(2n)_1$ part of the coset theory.
To make the discussion concrete,
we consider only the case of $n=4$, although
the case of $n \neq 4$ can be treated in the same way.

The building blocks of the construction are the permutation NIM-rep
\eqref{eq:npi} for $so(8)_1 \oplus so(8)_1$ and
the regular NIM-rep for $so(8)_2$,
from which one can obtain a NIM-rep \eqref{eq:finalpsi}
of the coset theory.
Since the spectrum of the permutation NIM-rep consists
of the representations of the form
$(\mu, \mu) \,(\mu \in \I_1 = P_+^{k=1}(so(8)))$,
the spectrum \eqref{eq:cosetspec} of the Ishibashi states
reads
\beq
  \hat{\E} = \{(\Lambda_0, \Lambda_0; \nu) \in \hat{\I}\}
  = \{u_+, u_-, \phi_+, \phi_-, \chi_2 \} .
\label{eq:Epi_temp}
\eeq
We therefore obtain five Ishibashi states.
The Cardy states of the permutation NIM-rep is also labeled by
$\alpha \in \I_1$.
Applying the formula \eqref{eq:cosetboundary}, one obtains
\beq
  \hat{\V} = \{(\alpha; \beta) \, | \,
        \alpha \in \I_1, \beta \in \I_2, \,
        b_\alpha(J) = b_\beta(J), \,
        (J^2 \alpha; J \beta) = (\alpha; \beta) \,
        (\forall J \in \G) \} ,
\eeq
where we used the action of the simple currents \eqref{eq:sconVpi}
on the permutation NIM-rep.
Using $J^2 = 1$ for $so(8)$,
one obtains $\hat{\V}$ explicitly,
\beq
  \hat{\V} = \{
   (\Lambda_0; 2 \Lambda_0)_4 ,
   (\Lambda_0;   \Lambda_2)_1 ,
   (\Lambda_1; \Lambda_0 + \Lambda_1)_2 ,
   (\Lambda_3; \Lambda_0 + \Lambda_3)_2 ,
   (\Lambda_4; \Lambda_0 + \Lambda_4)_2 \}.
\eeq
The subscript in this equation stands for the length of
the brane identification orbit
\beq
  \{(\alpha; J \beta) \,|\, J \in \G\},
\eeq
which is equal to $4$ if there is no fixed point.
One can see that all the elements other than $(\Lambda_0; 2 \Lambda_0)$
are fixed points of the brane identification,
and we have to resolve them to obtain a consistent NIM-rep of
the coset theory.
The degeneracy of the fixed points are given by the ratio
of the orbit length to the order of the identification current group.
After resolving them, the number of the Cardy states
would therefore be
$1 + 4 + 2 \times 3 = 11$.
Since there are only five Ishibashi states \eqref{eq:Epi_temp} in $\hat{\E}$,
we have to find $11 - 5 = 6$ states besides those in $\hat{\E}$
for the resolution of the fixed points.
One can find these states by considering the action \eqref{eq:pihat}
of $\hat{\pi}$ on $\hat{\I}$.
Actually, one can see that not only the states in $\hat{\E}$ but
all the elements in $\hat{\I}$ are invariant under the action of
$\hat{\pi}$ due to the field identification.
Eventually, we obtain $6$ extra Ishibashi states to resolve the fixed
points in $\hat{\V}$.
The spectrum of the resolved NIM-rep
is however the same as that of the regular one.
In fact, one can show that
the diagonalization matrix \eqref{eq:finalpsi} obtained from the permutation
NIM-rep is considered to be the regular NIM-rep after
appropriately resolving the fixed points.

One can repeat the same analysis for the case of $n \neq 4$.
There are also the brane identification fixed points.
After the resolution of them, one again finds nothing new,
namely, the regular NIM-rep for even $n$ and
the $\hat{\omega}_2$-twisted NIM-rep for odd $n$.

\section{Relation with orbifold theories}
\label{sec:relation}

The diagonal coset theory
$so(2n)_1 \oplus so(2n)_1/so(2n)_2$
has the central charge $c= c_n^{k=1} + c_n^{k=1}-c_n^{k=2} = 1$, where
\begin{equation}
 c_n^k=\frac{(2n^2-n)k}{k+2n-2}
\end{equation}
is the central charge of the $SO(2n)$ WZW model at level $k$.
Therefore this model should belong to the class of rational CFTs
with $c=1$, which consists of
the circle theory $S^1$ with the radius $R=\sqrt{2n} \,(n\in \Z_{> 0})$,
\footnote{%
We set $\alpha' = 2$.}
its $\Z_2$ orbifold $S^1/\Z_2$ and
three exceptional models ${\bf T,O,I}$ \cite{Ginsparg}.
\footnote{We restrict ourselves to the case of
the diagonal modular invariants.}
Actually, by comparing the spectrum,
the $so(2n)$ diagonal coset theory is shown to be equivalent
with the orbifold theory at $r = \sqrt{n}$
\begin{equation}
 \frac{so(2n)_1 \oplus so(2n)_1}{so(2n)_2} \sim
  S^1(\sqrt{n})/\Z_2 ,
\end{equation}
where we introduced the normalized radius $r \equiv \frac{R}{\sqrt{2}}$
so that the self-dual point corresponds to $r=1$.

In this section, we use this equivalence of the $so(2n)$ coset theory
with the orbifold to check our calculation
for the twisted Cardy states in the coset theory.\footnote{%
Boundary states in the $c=1$ models have been studied
from the RCFT point of view also in \cite{Cappelli}.}
Namely, we shall find that both of the $\Z_2$ and the $\Z_3$ Cardy states
have their counterpart in the orbifold theory.
The $\Z_2$ states as well as the regular ones
are realized by the usual Neumann and Dirichlet states of the orbifold.
On the other hand, one can not find the $\Z_3$ states
within the Neumann and Dirichlet states; rather the $\Z_3$ states
correspond to the conformal boundary states of the orbifold,
which generalize those of $S^1$ constructed in
\cite{GRW,GR}.
(The construction of the conformal boundary states
for the orbifold (at $r=2$) is found in Appendix~\ref{sec:conformal}.)
This fact assures the validity of our calculation, in particular,
the mutual consistency of the twisted states with the regular ones.

\subsection{Regular Cardy states}

We begin with rewriting the regular Cardy states of the coset theory
in terms of the orbifold.
We review some basic facts about the boundary states in the orbifold
in Appendix~\ref{sec:DN}.

Since the Cardy states are expressed as linear combinations of
the Ishibashi states, we need to know
the expression of the Ishibashi states in the orbifold theory.
From the explicit form of the characters of the $so(2n)$ coset theory
(see Appendix~\ref{sec:bo}),
we can identify the corresponding Ishibashi states in
the orbifold at $r = \sqrt{n}$ as follows,
\bes
   \dket{u_{\pm}}    &=
     \frac{1}{2}\sum_{m\in \Z} (\dket{\DD(2n m)}\pm \dket{\NN(2m)} ),\\
   \dket{\phi_{\pm}} &=
     \frac{1}{2}\sum_{m\in \Z} (\dket{\DD(n(2m+1))}\pm i^n\dket{\NN(2m+1)}),\\
   \dket{\chi_s}     &=
     \frac{1}{\sqrt{2}}\sum_{m\in \Z} (\dket{\DD(2nm+s)}+\dket{\DD(2nm-s)}),\\
   \dket{\sigma_j}   &=
     \frac{1}{2\sqrt{2}}
     \Bigl[\dket{\DD_T(0)}+\dket{\NN_T(0)}
      +i^n(-1)^j (\dket{\DD_T(\pi)}+\dket{\NN_T(\pi)}) \Bigr],\\
   \dket{\tau_j}     &=
     \frac{1}{2\sqrt{2}}
     \Bigl[\dket{\DD_T(0)}-\dket{\NN_T(0)}
      +i^n(-1)^j (\dket{\DD_T(\pi)}-\dket{\NN_T(\pi)}) \Bigr] .
\label{eq:Ishibashi_relation}
\ees
See Appendix~\ref{sec:DN}
for our notation of the Ishibashi states in $S^1/\Z_2$.
Then it is straightforward to express the regular Cardy states
$\ket{\mu} = \sum_{\nu \in \hat{\I}} \hat{S}_{\mu\nu}
\frac{1}{\sqrt{\hat{S}_{0\nu}}} \dket{\nu}$ in terms of the boundary states
of the orbifold \cite{Cappelli},
\bes
   \ket{u_\pm}    &= \ket{\DD[0];\pm}_\mathcal{O}, \quad
   \ket{\phi_\pm}  = \ket{\DD[\pi];\pm}_\mathcal{O}, \quad
   \ket{\chi_s}    = \ket{\DD[\tfrac{s}{n}\pi]}_\mathcal{O}, \\
   \ket{\sigma_j} &= \ket{\NN[j\pi];+}_\mathcal{O},\quad
   \ket{\tau_j}    = \ket{\NN[j\pi];-}_\mathcal{O}.
\label{eq:Cardy_relation}
\ees
To be precise, we can multiply each Ishibashi state of the orbifold
by some phases in our dictionary \eqref{eq:Ishibashi_relation}
without changing the normalization.
We have fixed this ambiguity by requiring that
the regular state $\ket{u_+}$ corresponds to one of the fractional
D-particle $\ket{\DD[0];+}$ sitting at $x=0$.

\subsection{Twisted Cardy states}

\subsubsection{$\Z_2$ states}

We can identify the $\Z_2$ twisted Cardy states of
the $so(2n)$ coset theory with the boundary states
of the orbifold in the same way as the regular ones.
To obtain the expression
of the $\Z_2$ twisted Ishibashi states in the orbifold theory,
we have two types of conditions.
The first one is that
the overlap of the twisted states with themselves should yield
the ordinary character of the coset theory,
which is the same condition as we used in deriving
eq.\eqref{eq:Ishibashi_relation}.
The second is that the overlap with the regular states should be
equal to the twining character for the $\Z_2$ automorphism,
which is the branching function of $A_{2n-3}^{(2)}$.
See Appendix~\ref{sec:bt} for their explicit form.
The $\Z_2$ twisted Ishibashi states can be determined from these
two conditions together with our expression \eqref{eq:Ishibashi_relation}
for the regular states.
The result is as follows,
\bes
    \dket{u_\pm}^{\Z_2}  &=
    \frac{1}{2} \sum_{m\in \Z} (-1)^m (\dket{\DD(2nm)}\pm\dket{\NN(2m)}) ,\\
    \dket{\chi_s}^{\Z_2} &=
    \frac{1}{\sqrt{2}} \sum_{m\in \Z} (-1)^m
      (\dket{\DD(2nm+s)}+\dket{\DD(2nm-s)}) .
\ees
Then one finds that
the $\Z_2$ twisted Cardy states \eqref{eq:psihat_omega2}
can be written in the form
\beq
  \ket{(\tilde{\Lambda}_0,\tilde{\Lambda}_0;\tilde{\lambda}_s)} =
        \ket{\DD[\tfrac{1}{n}(s + \tfrac{1}{2})\pi]}_\mathcal{O}, \quad
  \ket{(\tilde{\Lambda}_0,\tilde{\Lambda}_{n-1};
        \tilde{\Lambda}_0+\tilde{\Lambda}_{n-1})} =
        \ket{\NN[\tfrac{\pi}{2}]}_\mathcal{O}.
\eeq
Since the right-hand side of these equations coincide with
the usual boundary states of the orbifold,
the mutual consistency of the $\Z_2$ states with the regular
ones is manifest.

\subsubsection{$\Z_3$ states}

For $n = 4$, we have the $\Z_3$ twisted Cardy states.
As we will see shortly, one can not express the $\Z_3$ states
within the usual Neumann and Dirichlet states of the orbifold
at $r = \sqrt{n} = 2$, and we are led to introduce the conformal
boundary states of the orbifold in order to express the $\Z_3$ states.

As we have seen in the last section,
there are two $\Z_3$ Cardy states \eqref{eq:psihat_omega3}.
For convenience, we introduce a short-hand notation for them,
\beq
  \ket{0}_{\Z_3} =
  \ket{(\tilde{\Lambda}_0, \tilde{\Lambda}_0; 2\tilde{\Lambda}_0)} , \quad
  \ket{1}_{\Z_3} =
  \ket{(\tilde{\Lambda}_0, \tilde{\Lambda}_0; \tilde{\Lambda}_1)} .
  \label{eq:Z3short}
\eeq
Let us consider
the overlap of these states with the regular state $\ket{u_\pm}$,
\bes
  \bra{u_\pm} \tilde{q}^{H_c} \ket{0}_{\Z_3} &=
  \frac{1}{\eta(q)} \sum_{N \in \Z} q^{(N+\frac{1}{6})^2} , \\
  \bra{u_\pm} \tilde{q}^{H_c} \ket{1}_{\Z_3} &=
  \frac{1}{\eta(q)} \sum_{N \in \Z} q^{(N+\frac{1}{3})^2} ,
\label{eq:Z3overlap}
\ees
which are readily obtained from the explicit form of these states.
From this expression, one can conclude that
the $\Z_3$ states are not realized by either the Dirichlet or the Neumann
states.
In order to see this,
note that the spectrum \eqref{eq:Ehat_omega3} of
the $\hat{\omega}_3$-invariant Ishibashi states consists of two states
$u_+, \chi_2$.
The orbifold expression \eqref{eq:Ishibashi_relation} of the regular
Ishibashi states shows that these two states $u_+, \chi_2$,
do not contain the fields in the twisted sectors of the orbifold.
Therefore the $\Z_3$ Cardy states \eqref{eq:Z3short} should consist of
only the fields in the untwisted sector, and
only the untwisted sector part of $\ket{u_\pm}$,
which is given by $\frac{1}{\sqrt{2}}\ket{\DD[0]}$,
contributes in the overlap \eqref{eq:Z3overlap}.
However,
from Appendix~\ref{sec:DN}, we find that any states of the Dirichlet
or the Neumann types do not have the overlap
of the form \eqref{eq:Z3overlap} with $\frac{1}{\sqrt{2}}\ket{\DD[0]}$.
Actually, the overlap of $\ket{u_+} = \frac{1}{\sqrt{2}}\ket{\DD[0]} + \cdots$
with non-fractional states takes the form
\bes
  \bra{u_+} \tilde{q}^{H_c} \ket{\DD[x]}_\mathcal{O} &=
  \frac{1}{\eta(q)} \sum_{N \in \Z} q^{4(N+\frac{x}{2\pi})^2} , \\
  \bra{u_+} \tilde{q}^{H_c} \ket{\NN[\theta]}_\mathcal{O} &=
  \frac{1}{\eta(q)} \sum_{N \in \Z} q^{(N+\frac{1}{4})^2} ,
\ees
which is clearly different from that in \eqref{eq:Z3overlap}.
Therefore we can conclude that the $\Z_3$ states \eqref{eq:Z3short}
are not realized within the usual states of the orbifold,
which preserve the chiral algebra $u(1)/\Z_2$ of $S^1/\Z_2$.
We need states other than those keeping $u(1)/\Z_2$ for
describing the $\Z_3$ Cardy states in the orbifold.
Actually, in the orbifold, there exist many boundary states
besides those of the Neumann and the Dirichlet types.
This can be understood from the fact that,
in order to obtain a boundary CFT, it may be enough to require
that only the Virasoro algebra, rather than its extension,
is preserved at the boundaries.
In general, finding the conformal boundary states
that preserve only the Virasoro algebra is a difficult task
since a CFT is irrational with respect to the Virasoro algebra for $c>1$.
For $c=1$, however, this problem is within the reach of the known techniques
and has been solved for the case of $S^1$ \cite{GRW,GR}
(see also \cite{Janik,Tseng}).
This construction of the conformal boundary states
is readily generalized to the orbifold $S^1/\Z_2$.
We report the detail of the construction in Appendix~\ref{sec:conformal}
for the case of $r=2$.

The construction of the conformal boundary states is
based on the fact that the CFT for $S^1$ at $r=1$ is equivalent
with the $SU(2)$ WZW model at level 1.
The conformal boundary states of $S^1$ at $r=1$ are therefore
parametrized by $SU(2)$ \cite{GRW,GR}.
Since $S^1/\Z_2$ at $r=n \in \Z_{\ge 1}$ can be considered
as an orbifold of the $SU(2)$ WZW model by the discrete subgroup
$\Gamma = D_n$ of $SU(2)$,\footnote{%
To be precise, the dihedral group $D_n$ is a subgroup of
$SO(3) \cong SU(2)/\{1,-1\}$, where $\{1, -1\}$ is the center of
$SU(2)$.
As a subgroup of $SU(2)$, we should consider
the binary dihedral group $\mathcal{D}_n$.}
the conformal boundary states of $S^1/\Z_2$ are also parametrized
by $g \in SU(2)$, which we denote by $\ket{g}_\Gamma$
(see Appendix~\ref{sec:conformal} for detail).
One can find the $\Z_3$ Cardy states \eqref{eq:Z3short} within these
conformal boundary states of $S^1/\Z_2$.
The result is as follows,\footnote{%
One can also realize the $\Z_3$ states by setting
$g=\frac{1}{2}(1-\gamma_1-\gamma_2-\gamma_3)$.
These two choices for $g$ correspond to the twist 
by $\hat{\omega}_3$ and by $(\hat{\omega}_3)^2$.}
\bes
  &\ket{0}_{\Z_3} = \ket{g}_\Gamma , \quad
  \ket{1}_{\Z_3} = \ket{-g}_\Gamma ,  \\
  &g= \frac{1}{2} ( 1 + \gamma_1 + \gamma_2 + \gamma_3) =
    \frac{1}{2} \begin{pmatrix}
      1 + i & 1+i \\
      -1 +i  & 1-i  \end{pmatrix} ,
\label{eq:Z3answer}
\ees
where $\gamma_a = i\sigma_a$ form the dihedral group
$D_2 = \{1, \gamma_1, \gamma_2, \gamma_3 \} \cong \Z_2 \times \Z_2$.
With this choice, one can show that the overlap \eqref{eq:Z3overlap}
is reproduced by using the formula
\eqref{eq:conformal_overlap} in Appendix~\ref{sec:conformal}.
Moreover, the overlap with the $\Z_3$ states themselves
\bes
  {}_{\Z_3}\bra{0} \tilde{q}^{H_c} \ket{0}_{\Z_3} &=
  {}_{\Z_3}\bra{1} \tilde{q}^{H_c} \ket{1}_{\Z_3}  =
  \frac{1}{\eta(q)}\left( \sum_{N \in \Z} q^{N^2} +
                   3 \sum_{N \in \Z} q^{(N + \frac{1}{4})^2} \right) , \\
  {}_{\Z_3}\bra{0} \tilde{q}^{H_c} \ket{1}_{\Z_3}  &=
  \frac{1}{\eta(q)}\left( \sum_{N \in \Z} q^{(N+\frac{1}{2})^2} +
                   3 \sum_{N \in \Z} q^{(N + \frac{1}{4})^2} \right) ,
\ees
can also be reproduced from our solution \eqref{eq:Z3answer}.
Hence we can conclude that the $\Z_3$ Cardy states of the $so(8)$ coset theory
are realized by the conformal boundary states of the orbifold.
This fact again shows the validity of our argument about the mutual consistency
of the Cardy states.

\section{Discussions}

In this paper, we have argued the mutual consistency of the twisted
boundary states with the untwisted ones in coset CFTs.
We have constructed explicitly 
the twisted Cardy states of the coset theory and 
have shown that they have well-defined overlaps with the untwisted
states. 
We have also pointed out that the overlap of the twisted states
can be described in terms of the branching functions
of twisted affine Lie algebras. 
As a check of our argument,
we have given a detailed study of the twisted boundary states in
the $so(2n)$ diagonal coset theory
and have confirmed
their consistency through the equivalence with the orbifold $S^1/\Z_2$. 
As a by-product, we have found a new type of boundary states in
$S^1/\Z_2$, which break $u(1)/\Z_2$ but 
preserve the Virasoro algebra.

A natural problem related with our result is the issue of the mutual
consistency of NIM-reps other than those originating from
the automorphisms of the chiral algebra. 
Actually, there exists many non-trivial NIM-reps for $so(2n)$ 
at level 2 \cite{Gannon} and
we can construct the corresponding NIM-reps of the diagonal coset theory
combining them with the NIM-rep of $so(2n)_1$.
It is interesting to see whether these NIM-reps are non-trivial or not.
If non-trivial, it would be
worth studying about their consistency with the regular one. 

Our construction of the twisted Cardy states for the triality
of $so(8)$ exhibits that 
the technique of RCFTs can be used for studying the conformal boundary
states in the $c=1$ models.
This may be extended to the case of $c > 1$ by considering
$su(n)_1$ and its realization by a $(n-1)$-dimensional torus $T^{n-1}$.
The role of the Virasoro algebra is taken over by the $W_n$ algebra
and we should consider the $W_n$ boundary states instead of the conformal
ones.
It would be interesting to find an appropriate realization of $T^{n-1}$
as a RCFT equipped with automorphisms and observe 
whether the corresponding twisted boundary states yield the $W_n$ states.

\vskip \baselineskip
\noindent
\textbf{Acknowledgement: }

We would like to thank H.~Awata, M.~Kato, K.~Ito and T.~Tani
for helpful discussions.

\bigskip
\appendix
\section{Modular transformation matrices of $D_n^{(r)}$}
\label{sec:S}

\subsection{$D_n^{(1)}$}
\label{sec:S1}

\noindent
\underline{$k=1$}

\noindent
There are four integrable representations for $D_{n, 1}^{(1)} = so(2n)_1$,
\beq
  \I_1 = P_+^{k=1}(so(2n))
       = \{\Lambda_0, \Lambda_1, \Lambda_{n-1}, \Lambda_n \} .
\eeq
$\Lambda_j$ is the $j$-th fundamental weight of $D_n^{(1)}$.
The modular transformation matrix reads
\beq
  \frac{1}{2} \times \quad
  \begin{array}{c | cccc}
              & \Lambda_0 & \Lambda_1 & \Lambda_{n-1} & \Lambda_n \\ \hline
   \Lambda_0     & 1 &  1 &       1 &       1 \\
   \Lambda_1     & 1 &  1 &      -1 &      -1 \\
   \Lambda_{n-1} & 1 & -1 &  (-i)^n & -(-i)^n \\
   \Lambda_n     & 1 & -1 & -(-i)^n &  (-i)^n \\[2\jot] \hline\hline
   \text{class}  & O & V  & C & S \rule{0cm}{5mm}
  \end{array}
\label{eq:Dnk1}
\eeq
where the last row shows the conjugacy class of each representation.

\bigskip
\noindent
\underline{$k=2$}

\noindent
There are $n + 7$ integrable representations for $so(2n)_2$,
\bes
  \I_2  &= \{2\Lambda_0, 2\Lambda_1, 2\Lambda_{n-1}, 2\Lambda_n, \\
        & \quad\quad
        \Lambda_0 + \Lambda_{n-1},  \Lambda_0 + \Lambda_{n},
        \Lambda_1 + \Lambda_{n-1},  \Lambda_1 + \Lambda_{n},
        \lambda_j \,(j=1,2,\dots,n-1) \},
\ees
where $\lambda_j$ are defined as
\beq
\lambda_j=
   \begin{cases}
      \Lambda_0+\Lambda_1 & (j=1) ,\\
      \Lambda_j&(2\leq j \leq n-2) ,\\
      \Lambda_{n-1}+\Lambda_{n} & (j=n-1) .
   \end{cases}
\eeq
The modular transformation matrix reads
\bea
  &\frac{1}{\sqrt{8n}} \times \notag\\
  &{\scriptsize
  \renewcommand{\arraystretch}{2}
  \begin{array}{c | ccccccccc}
              & 2\Lambda_0 & 2\Lambda_1 & 2\Lambda_{n-1} & 2\Lambda_n &
                \lambda_s  &
                \Lambda_0 + \Lambda_{n-1} & \Lambda_0 + \Lambda_{n} &
                \Lambda_1 + \Lambda_{n} & \Lambda_1 + \Lambda_{n-1} \\
  \hline
   2\Lambda_0     & 1 &  1 &  1 & 1 & 2 &
                  \sqrt{n} & \sqrt{n} & \sqrt{n} & \sqrt{n} \\
   2\Lambda_1     & 1 &  1 &  1 & 1 & 2 &
                  -\sqrt{n} & -\sqrt{n} & -\sqrt{n} & -\sqrt{n} \\
   2\Lambda_{n-1} & 1 &  1 &  (-1)^n & (-1)^n & 2 (-1)^s &
                  (-i)^n \sqrt{n} & -(-i)^n \sqrt{n} &
                  (-i)^n \sqrt{n} & -(-i)^n \sqrt{n} \\
   2\Lambda_n     & 1 &  1 &  (-1)^n & (-1)^n & 2 (-1)^s &
                  -(-i)^n \sqrt{n} & (-i)^n \sqrt{n} &
                  -(-i)^n \sqrt{n} & (-i)^n \sqrt{n} \\
   \lambda_r      & 2 &  2 &  2(-1)^r & 2(-1)^r & 4 \cos \tfrac{\pi rs}{n} &
                    0 &  0 & 0 & 0 \\
   \Lambda_0 + \Lambda_{n-1} &
             \sqrt{n} & -\sqrt{n} & (-i)^n\sqrt{n} &  -(-i)^n \sqrt{n} & 0 &
              a & b & -a & -b \\
   \Lambda_0 + \Lambda_{n} &
             \sqrt{n} & -\sqrt{n} & -(-i)^n\sqrt{n} &  (-i)^n \sqrt{n} & 0 &
              b & a & -b & -a \\
   \Lambda_1 + \Lambda_{n} &
             \sqrt{n} & -\sqrt{n} & (-i)^n\sqrt{n} &  -(-i)^n \sqrt{n} & 0 &
              -a & -b & a & b \\
   \Lambda_1 + \Lambda_{n-1} &
             \sqrt{n} & -\sqrt{n} & -(-i)^n\sqrt{n} &  (-i)^n \sqrt{n} & 0 &
              -b & -a & b & a \\[2\jot] \hline\hline
   \text{class} & O & O &
                \begin{matrix} O \text{($n$:even)} \\
                               V \text{($n$: odd)} \end{matrix} &
                \begin{matrix} O \text{($n$:even)} \\
                               V \text{($n$: odd)} \end{matrix} &
                \begin{matrix} O \text{($s$:even)} \\
                               V \text{($s$: odd)} \end{matrix} &
                C & S & C & S
  \end{array}
  } \notag\\[2\jot]
  &\text{where}\quad a = \sqrt{\frac{n}{2}}(1 + (-i)^n) , \quad
                     b = \sqrt{\frac{n}{2}}(1 - (-i)^n) .
\eea

\bigskip
\noindent
\underline{simple currents}

\noindent
For even $n$, the simple current group $\G$ of $so(2n)_k$
is $\Z_2 \times \Z_2$,
whose action on the fundamental weights reads
\bes
  &J_V : (\Lambda_0, \Lambda_1, \Lambda_2, \dots,
          \Lambda_{n-2}, \Lambda_{n-1}, \Lambda_{n}) \mapsto
         (\Lambda_1, \Lambda_0, \Lambda_2, \dots,
          \Lambda_{n-2}, \Lambda_{n}, \Lambda_{n-1}) , \\
  &J_S : (\Lambda_0, \Lambda_1, \Lambda_2, \dots,
          \Lambda_{n-2}, \Lambda_{n-1}, \Lambda_{n}) \mapsto
         (\Lambda_n, \Lambda_{n-1}, \Lambda_{n-2}, \dots,
          \Lambda_{2}, \Lambda_{1}, \Lambda_{0}) , \\
  &J_C = J_V J_S , \quad (J_V)^2 = (J_S)^2 = 1 .
\ees
For odd $n$, $\G$ is $\Z_4$ and generated by $J_S$,
\bes
  &J_S : (\Lambda_0, \Lambda_1, \Lambda_2, \dots,
          \Lambda_{n-2}, \Lambda_{n-1}, \Lambda_{n}) \mapsto
         (\Lambda_n, \Lambda_{n-1}, \Lambda_{n-2}, \dots,
          \Lambda_{2}, \Lambda_{0}, \Lambda_{1}) , \\
  &J_V = (J_S)^2 , \quad J_C = (J_S)^3 , \quad (J_S)^4 = 1 .
\ees
The monodromy charge $b_\mu(J)$ for $J \in \G$ is a function of
the conjugacy class of $\mu$ and written as
\beq
  b_\mu(J) = \quad
  \begin{array}{r | rccc}
              & \mu = O & V & C & S \\ \hline
   J = J_O     & 1 &  1 &       1 &       1 \\
   J_V     & 1 &  1 &      -1 &      -1 \\
   J_C     & 1 & -1 &  (-i)^n & -(-i)^n \\
   J_S     & 1 & -1 & -(-i)^n &  (-i)^n
  \end{array}
\label{eq:bD1}
\eeq
which is nothing but the character table of $\G$.

\bigskip
\subsection{$D_{n}^{(2)}$}
\label{sec:S2}

The modular transformation of the characters of
the twisted affine Lie algebra $D_{n}^{(2)}$ gives rise to
those of $A_{2n-3}^{(2)}$, and vice versa.
Since the twining characters of $so(2n)_k$ for the $\Z_2$ automorphism
$\omega_2$ coincide with the ordinary characters
of $A_{2n-3}^{(2)}$ at level $k$,
we can label the integrable representations of $A_{2n-3}^{(2)}$
by the $\omega_2$-invariant representations of $so(2n)_k$,
which form the set $\I^{\omega_2}_k \subset \I_k$.
We therefore use $\I^{\omega_2}_k$ instead of $P_+^k(A_{2n-3}^{(2)})$
and regard the modular transformation matrix $\tilde{S}$
for $(D_{n}^{(2)},A_{2n-3}^{(2)})$
as a matrix between $\tilde{\I}_k = P_+^k(D_{n}^{(2)})$
and $\I^{{\omega_2}}_k$.

\medskip
\noindent
\underline{$k=1$}

\noindent
There are two integrable representations for $k=1$,
\bes
  \tilde{\I}_1 &= \{\tilde{\Lambda}_0, \tilde{\Lambda}_{n-1} \}, \\
  \I^{\omega_2}_1  &= \{\mu \in \I_1 \, | \, {\omega_2} \mu = \mu \} =
  \{ \Lambda_0, \Lambda_1 \} ,
\ees
where
$\tilde{\Lambda}_j \,(j = 0,1,\dots,n-1)$
are the fundamental weights of $D_{n}^{(2)}$.
The modular transformation matrix reads
\beq
  \frac{1}{\sqrt{2}} \times \quad
{\renewcommand{\arraystretch}{1.2}
  \begin{array}{c | cc}
              & \Lambda_0 & \Lambda_1 \\ \hline
   \tilde{\Lambda}_0         & 1 &   1  \\
   \tilde{\Lambda}_{n-1}     & 1 &  -1
  \end{array} }
\eeq

\noindent
\underline{$k=2$}

\noindent
There are $n+1$ integrable representations for $k=2$,
\bes
  \tilde{\I}_2 &= \{\tilde{\lambda}_j \, (j = 0,1,\dots,n-1),
  \tilde{\Lambda}_0 + \tilde{\Lambda}_{n-1} \}, \\
  \I^{\omega_2}_2  &= \{\mu \in \I_2 \, | \, {\omega_2} \mu = \mu \} =
  \{ 2\Lambda_0, 2\Lambda_1, \lambda_j \, (j = 1,2,\dots,n-1) \} ,
\ees
where $\tilde{\lambda}_j$ are defined as
\begin{equation}
 \tilde{\lambda_j} =
    \begin{cases}
       2\tilde{\Lambda}_0     & j=0 , \\
        \tilde{\Lambda}_j     & 1 \leq j \leq n-2 , \\
       2\tilde{\Lambda}_{n-1} & j=n-1 .
    \end{cases}
\label{eq:lambdatilde}
\end{equation}
The modular transformation matrix reads
\beq
  \frac{1}{\sqrt{2n}} \times \quad
{\renewcommand{\arraystretch}{1.5}
  \begin{array}{c | ccc}
              & 2\Lambda_0 & 2\Lambda_1 & \lambda_s \\ \hline
   \tilde{\lambda}_r   & 1 &  1  &
                         2 \cos \bigl(\frac{\pi}{2n}(2r + 1)s\bigr) \\
   \tilde{\Lambda}_0 + \tilde{\Lambda}_{n-1}     & \sqrt{n} &  -\sqrt{n} & 0
  \end{array} }
\eeq

\noindent
\underline{simple currents}

\noindent
For the modular transformation matrix
$\tilde{S}_{\tilde{\mu} \nu} \,
(\tilde{\mu} \in \tilde{\I}_k, \nu \in \I^{\omega_2}_k)$,
we have two types of actions of simple currents
since two sets, $\tilde{\I}_k$ and $\I^{\omega_2}_k$, are distinct.
The action on $\tilde{\I}_k$ is caused by the diagram automorphism
of $D_{n}^{(2)}$ (see Fig.~\ref{fig:auto}),
which we denote by $\tilde{J}$,
\beq
  \tilde{J} : \tilde{\Lambda}_j \mapsto \tilde{\Lambda}_{n-1-j} , \quad
  \tilde{J}^2 = 1 .
\eeq
The action on $\I^{\omega_2}_k$ originates from the diagram automorphism
of $A_{2n-3}^{(2)}$ and is of order $2$.
Since $\I^{\omega_2}_k \subset \I_k$, this action can be realized
by an element of the simple current group $\G$ for $D_n^{(1)}$.
Among four elements of $\G$, only $J_V$ (besides $J_O = 1$)
leaves $\I^{\omega_2}_k$ invariant.
The action on $\I^{\omega_2}_k$ is therefore caused by $J_V$.
Under these actions of simple currents,
$\tilde{S}_{\tilde{\mu} \nu}$ transforms as follows \cite{Ishikawa}
\begin{subequations}
\bea
  \tilde{S}_{\tilde{J}\!\tilde{\mu}\, \nu} &=
  \tilde{S}_{\tilde{\mu} \nu} b_\nu(J_S) , \\
  \tilde{S}_{\tilde{\mu}\,J_V\!\nu} &=
  \tilde{b}_{\tilde{\mu}} \tilde{S}_{\tilde{\mu} \nu} ,
\eea
\end{subequations}
where $b_\nu(J_S)$ is the monodromy charge \eqref{eq:bD1} for $so(2n)$
and $\tilde{b}_{\tilde{\mu}}$ is defined as
\beq
  \tilde{b}_{\tilde{\mu}} = (-1)^{\tilde{\mu}_{n-1}} .
\eeq
Here $\tilde{\mu}_{n-1}$ is the Dynkin label of $\tilde{\mu}$.
For example,
$\tilde{b}_{\tilde{\mu}} = -1$ for only
$\tilde{\Lambda}_{n-1}$ in $\tilde{\I}_1$
and
$\tilde{\Lambda}_0 + \tilde{\Lambda}_{n-1}$ in $\tilde{\I}_2$.

\bigskip
\subsection{$D_{4}^{(3)}$}
\label{sec:S3}

For $so(8)$, besides the $\Z_2$ automorphism ${\omega_2}$,
we have an automorphism $\omega_3$ of order $3$
which yields the twisted affine Lie algebra $D_4^{(3)}$.
In the same way as the $\Z_2$ case,
we can label the integrable representations of $D_4^{(3)}$
also by the $\omega_3$-invariant representations of $so(8)_k$,
which form the set $\I^{\omega_3}_k \subset \I_k$.
Namely,
we have an isomorphism
$\I^{\omega_3}_k \cong \tilde{\I}^3_k = P_+^k(D_4^{(3)})$.
Although the modular transformation of the characters of
$D_{4}^{(3)}$ gives rise to those of $D_{4}^{(3)}$ itself,
it is convenient for describing the twisted Cardy states
to regard the modular transformation matrix $\tilde{S}$ for $D_4^{(3)}$
as a matrix between $\tilde{\I}^3_k$ and $\I^{\omega_3}_k$.

\medskip
\noindent
\underline{$k=1$}

\noindent
There is only one integrable representation for $k=1$,
\bes
  \tilde{\I}^3_1   &= \{\tilde{\Lambda}_0  \}, \\
  \I^{\omega_3}_1  &= \{\mu \in \I_1 \, | \, \omega_3 \mu = \mu \} =
  \{ \Lambda_0 \} ,
\ees
where
$\tilde{\Lambda}_j \,(j = 0,1,2)$
are the fundamental weights of $D_4^{(3)}$.

\noindent
\underline{$k=2$}

\noindent
There are two integrable representations for $k=2$,
\bes
  \tilde{\I}^3_2   &= \{ 2\tilde{\Lambda}_0, \tilde{\Lambda}_1 \} , \\
  \I^{\omega_3}_2  &= \{\mu \in \I_2 \, | \, \omega \mu = \mu \} =
  \{ 2\Lambda_0, \Lambda_2 \} .
\ees
The modular transformation matrix reads
\beq
  \frac{1}{\sqrt{2}} \times \quad
{\renewcommand{\arraystretch}{1.5}
  \begin{array}{c | cc}
              & 2\Lambda_0 & \Lambda_2  \\ \hline
   2\tilde{\Lambda}_0   & 1 &  1  \\
    \tilde{\Lambda}_1   & 1 &  -1
  \end{array} }
\eeq
\medskip
Since there is no diagram automorphism for $D_4^{(3)}$,
the simple current group for $\tilde{\I}^3_k$ is trivial.

\bigskip
\section{Characters of
$so(2n)_1 \oplus so(2n)_1/so(2n)_2$}
\label{sec:branching}

\subsection{Ordinary characters}
\label{sec:bo}

There are $n+7$ independent representations
in the branching of $so(2n)_1 \oplus so(2n)_1$ into $so(2n)_2$
as is shown in Table~\ref{tab:field}.
The corresponding branching functions are
determined by decomposing the product of
characters of $so(2n)_1$ into those of $so(2n)_2$
\cite{KW} (see \S 13.15 in \cite{Kac}):
\bes
  u_\pm(q) &= \frac{1}{\eta(q)} \frac{1}{2} \biggl(
  \sum_{N \in \Z} q^{n N^2} \pm \sum_{N \in \Z} (-1)^N q^{N^2} \biggr), \\
  \phi_\pm(q) &= \frac{1}{\eta(q)} \frac{1}{2}
  \sum_{N \in \Z} q^{n (N+\frac{1}{2})^2}, \\
  \chi_r(q) &= \frac{1}{\eta(q)}
  \sum_{N \in \Z} q^{n (N+\frac{r}{2n})^2} \quad (r = 1, 2,\dots,n-1), \\
  \sigma_j(q) &= \frac{1}{\eta(q)}
  \sum_{N \in \Z} q^{4 (N+\frac{1}{8})^2} \quad (j = 0,1), \\
  \tau_j(q) &= \frac{1}{\eta(q)}
  \sum_{N \in \Z} q^{4 (N+\frac{3}{8})^2} \quad (j = 0,1) .
\label{eq:D1branching}
\ees
Here
we denote the branching functions by the same symbol as the corresponding
representations, \textit{e.g.},
$u_+(q)$ stands for $\chi_{(\Lambda_0,\Lambda_0; 2\Lambda_0)}$.

\subsection{Twining characters}
\label{sec:bt}

\subsubsection{$\Z_2$-automorphism}

The $\Z_2$-automorphism ${\omega_2}$ of $so(2n)$ yields
an automorphism $\hat{\omega}_2$ of the diagonal coset
$so(2n)_1 \oplus so(2n)_1/so(2n)_2$,
which acts on the representations as
\beq
  \hat{\omega}_2 : (\mu_1, \mu_2; \nu) \mapsto
                 ({\omega_2} \mu_1, {\omega_2} \mu_2; {\omega_2} \nu) .
\eeq
There are $n+1$ representations invariant under $\hat{\omega}_2$,
which form the set $\hat{\I}^{\hat{\omega}_2}$,
\beq
  \hat{\I}^{\hat{\omega}_2} = \{u_\pm, \chi_r \, (r = 1,2,\dots,n-1)\}
  \subset \hat{\I} .
\eeq
Note that all the elements of $\hat{\I}^{\hat{\omega}_2}$ can be written
in the form
$(\Lambda_0, \mu; \nu), {\omega_2} \mu = \mu, {\omega_2} \nu = \nu$.
Hence the twining characters for the coset theory
can be determined from the following equations
\beq
  \chi^{\omega_2}_{\Lambda_0} \chi^{\omega_2}_{\mu} =
  \sum_\nu \chi^{\hat{\omega}_2}_{(\Lambda_0, \mu; \nu)} \chi^{\omega_2}_\nu ,
\eeq
where $\chi^{\omega_2}_\mu$ is the twining character of $so(2n)$.
Since $\chi^{\omega_2}_\mu$ coincides
with the ordinary character of $A_{2n-3}^{(2)}$,
one can regard the above equation as the branching rule of
$A_{2n-3, 2}^{(2)} \subset A_{2n-3, 1}^{(2)} \oplus A_{2n-3, 1}^{(2)}$
and $\chi^{\hat{\omega}_2}_{(\Lambda_0, \mu; \nu)}$
as the corresponding branching function,
which can be obtained in the same way as $D_n^{(1)}$.
The result is written as follows,
\bes
  u^{\Z_2}_\pm(q)  &\equiv \chi^{\hat{\omega}_2}_{u_\pm}(q) =
  \frac{1}{\eta(q)} \frac{1}{2} \biggl(
  \sum_{N \in \Z} (-1)^N q^{n N^2}
  \pm \sum_{N \in \Z} (-1)^N q^{N^2} \biggr), \\
  \chi^{\Z_2}_r(q) &\equiv \chi^{\hat{\omega}_2}_{\chi_r}(q) =
  \frac{1}{\eta(q)}
  \sum_{N \in \Z} (-1)^N q^{n (N+\frac{r}{2n})^2} \quad (r = 1, 2,\dots,n-1).
\label{eq:D2branching}
\ees

\subsubsection{$\Z_3$-automorphism}

The case of the $\Z_3$-automorphism $\omega_3$ for $so(8)$
can be treated in exactly the same way as the $\Z_2$ case
discussed above.
We denote by $\hat{\omega}_3$ the associated automorphism
of the $so(8)$ diagonal coset theory.
Among $11$ representations in $\hat{\I}$,
only two of them are invariant under $\hat{\omega}_3$,
\beq
  \hat{\I}^{\hat{\omega}_3} = \{u_+, \chi_2 \}
  \subset \hat{\I} .
\eeq
The corresponding twining characters of the coset theory
can be obtained as the branching function of
$D_{4,2}^{(3)} \subset D_{4,1}^{(3)} \oplus D_{4,1}^{(3)}$.
The result is written as follows,
\bes
  u^{\Z_3}_{+}(q)  &\equiv \chi^{\hat{\omega}_3}_{u_+}(q) =
  \frac{1}{\eta(q)}
  \sum_{N \in \Z} e^{\frac{2\pi i}{3}N} q^{N^2} , \\
  \chi^{\Z_3}_2(q) &\equiv \chi^{\hat{\omega}_3}_{\chi_2}(q) =
  \frac{1}{\eta(q)}
  \sum_{N \in \Z + \frac{1}{2}} e^{\frac{2\pi i}{3}N} q^{N^2} .
\label{eq:D3branching}
\ees

\subsection{Characters of the twisted chiral algebra}
\label{sec:bt2}


The characters of the twisted chiral algebra
for the $\Z_2$-automorphism $\hat{\omega}_2$
is given by the branching functions of
$D_{n,2}^{(2)} \subset D_{n,1}^{(2)} \oplus D_{n,1}^{(2)}$,
which are determined by
\beq
  \chi_{\tilde{\Lambda}_0} \chi_{\tilde{\mu}} =
  \sum_{\tilde{\nu}} \chi_{(\tilde{\Lambda}_0, \tilde{\mu}; \tilde{\nu})}
                     \chi_{\tilde{\nu}} ,
\eeq
where
$\tilde{\mu} \in P_+^{k=1}(D_n^{(2)}), \tilde{\nu} \in P_+^{k=2}(D_n^{(2)})$.
There are $n+1$ branching functions and the result is written as follows,
\bes
  \chi_{(\tilde{\Lambda}_0, \tilde{\Lambda}_0; \tilde{\lambda}_j)}(q)
  & =  \frac{1}{\eta(q^2)}
  \sum_{N \in \Z} q^{2 n (N + \frac{1}{2n}(j + \frac{1}{2}))^2}
  \quad (j = 0,1,\dots, n-1) ,\\
  \chi_{(\tilde{\Lambda}_0, \tilde{\Lambda}_{n-1};
         \tilde{\Lambda}_0 + \tilde{\Lambda}_{n-1})}(q)
  & =  \frac{1}{\eta(q^2)}
  \sum_{N \in \Z} q^{2 (N + \frac{1}{4})^2} ,
\ees
where $\tilde{\lambda}_j$ are the integrable representations for $k=2$
defined in eq.\eqref{eq:lambdatilde}.

\bigskip
\section{Boundary states in the orbifold $S^1/\Z_2$}
\label{sec:DN}

\subsection{$S^1$}

The circle $S^1$ is expressed by a free boson $X$,
\footnote{We set $\alpha' = 2$.}
\begin{equation}
   X(z,\bar{z}) =
  x + i p_L \ln z + i p_R \ln \bar{z}
  +i \sum_{n\neq 0}\frac{1}{n} a_n z^{-n}
  +i \sum_{n\neq 0}\frac{1}{n} \tilde{a}_n {\bar z}^{-n}.
\end{equation}
Let $R$ be the radius of $S^1$.
We also introduce the normalized radius
\beq
  r \equiv \frac{R}{\sqrt{\alpha'}} = \frac{R}{\sqrt{2}},
\eeq
in which the self-dual point corresponds to $r=1$.
Then the spectrum of the left and right momenta $p_L,p_R,$ is written as
\bes
  p_L &= \frac{n}{R}+\frac{mR}{2}
       = \frac{1}{\sqrt{2}} \biggl(\frac{n}{r} + m r \biggr), \\
  p_R &= \frac{n}{R}-\frac{mR}{2}
       = \frac{1}{\sqrt{2}} \biggl(\frac{n}{r} - m r \biggr) ,
\label{eq:pspec}
\ees
where $n$ and $m$ stand for the momentum and
the winding number of the string, respectively.

The Dirichlet and Neumann boundary conditions are expressed on
the modes as\footnote{%
The zero modes are defined as $a_0=p_L$ and $\tilde{a}_0=p_R$.}
\begin{align}
 a_n+\tilde{a}_{-n}=0 &  \quad      (\text{Dirichlet}),\label{eq:bcD}\\
 a_n-\tilde{a}_{-n}=0 &  \quad      (\text{Neumann}).  \label{eq:bcN}
\end{align}
The Ishibashi states for these boundary conditions read
\begin{align}
  \dket{\DD(n)} &=
    \exp\left(\sum_{N>0}\frac{1}{N}a_{-N}\widetilde{a}_{-N}\right)
    \ket{(n,0)},\\
  \dket{\NN(m)} &=
    \exp\left(-\sum_{N>0}\frac{1}{N}a_{-N}\widetilde{a}_{-N}\right)
    \ket{(0,m)},
\end{align}
where $\ket{(n,m)}$ is the ground state with
momentum $n$ and winding number $m$.
The overlaps of these states take the form
\bea
  \dbra{\DD(n)} \tilde{q}^{H_c} \dket{\DD(n')} &=
    \delta_{n,n'} \frac{{\tilde q}^{\frac{n^2}{4 r^2}}}{\eta({\tilde q)}},\\
  \dbra{\NN(m)} \tilde{q}^{H_c} \dket{\NN(m')} &=
    \delta_{m,m'} \frac{{\tilde q}^{\frac{m^2 r^2}{4}}}{\eta({\tilde q)}},\\
  \dbra{\DD(n)} \tilde{q}^{H_c} \dket{\NN(m)} &=
    \delta_{n,0} \delta_{m,0} \frac{1}{\eta({\tilde q})}
    \sum_{N\in Z}(-1)^N {\tilde q}^{N^2},
\eea
where $\eta(q)$ is Dedekind's eta function and
$H_c = \frac{1}{2}(L_0 + \tilde{L}_0 - \frac{1}{12})$
is the closed string Hamiltonian.
Then one can write the Dirichlet and Neumann boundary states
in the form
\bea
  \ket{\DD[x]}      &= 2^{-1/4}\, r^{-1/2}
     \sum_{n \in \Z} e^{-i n x} \dket{\DD(n)}, \quad
     0 \le x < 2\pi , \\
  \ket{\NN[\theta]} &= 2^{-1/4}\, r^{1/2}
     \sum_{m \in \Z} e^{-i m \theta} \dket{\NN(m)}, \quad
     0 \le \theta < 2\pi .
\eea
$\ket{\DD[x]}$ corresponds to a D0-brane sitting at $X= x R$
while $\ket{\NN[\theta]}$ is a D1-brane
with the Wilson line $\theta$.
One can calculate the overlaps of these states from those of the Ishibashi
states,
\bea
  \bra{\DD[x]} \tilde{q}^{H_c} \ket{\DD[x']} &=
    \frac{1}{\eta(q)} \sum_{N \in \Z} q^{r^2(N + \frac{x-x'}{2\pi})^2} \\
  \bra{\NN[\theta]} \tilde{q}^{H_c} \ket{\NN[\theta']} &=
    \frac{1}{\eta(q)} \sum_{N \in \Z}
    q^{\frac{1}{r^2}(N + \frac{\theta-\theta'}{2\pi})^2} \\
  \bra{\DD[x]} \tilde{q}^{H_c} \ket{\NN[\theta]} &=
    \frac{1}{\eta(q)} \sum_{N \in \Z} q^{(N + \frac{1}{4})^2} .
\eea

\subsection{$S^1/\Z_2$}

The boundary states of the orbifold $S^1/\Z_2$ follow from
those of $S^1$ by the standard procedure,
\bea
  \ket{\DD[x]}_\mathcal{O} &=
    \frac{1}{\sqrt{2}}(\ket{\DD[x]}+\ket{\DD[-x]}),
    \quad x \neq 0, \,\pi, \\
  \ket{\NN[\theta]}_\mathcal{O} &=
    \frac{1}{\sqrt{2}}(\ket{\NN[\theta]}+\ket{\NN[-\theta]}),
    \quad \theta \neq 0, \,\pi .
\eea
The fixed points of the $\Z_2$ action $X \mapsto -X$
can be resolved taking into account the Ishibashi states
from the twisted sectors \cite{OA},
\bea
   \ket{\DD[0];\pm}_\mathcal{O}   &=
     \frac{1}{\sqrt{2}}\ket{\DD[0]} \pm 2^{-1/4} \dket{\DD_T(0)},\\
   \ket{\DD[\pi];\pm}_\mathcal{O} &=
     \frac{1}{\sqrt{2}}\ket{\DD[\pi]} \pm 2^{-1/4} \dket{\DD_T(\pi)},\\
   \ket{\NN[0];\pm}_\mathcal{O}   &=
     \frac{1}{\sqrt{2}}\ket{\NN[0]}
     \pm 2^{-1/4} \frac{1}{\sqrt{2}} (\dket{\NN_T(0)}+\dket{\NN_T(\pi)}),\\
   \ket{\NN[\pi];\pm}_\mathcal{O} &=
     \frac{1}{\sqrt{2}}\ket{\NN[\pi]}
     \pm 2^{-1/4} \frac{1}{\sqrt{2}} (\dket{\NN_T(0)}-\dket{\NN_T(\pi)}).
\eea
Here we introduced four Ishibashi states associated with
the fixed points $x = 0, \pi$,
\bea
   \dket{\DD_T(x)} &=
     \exp\! \left[
       \sum_{N = \frac{1}{2}, \frac{3}{2}, \dots}
       \frac{1}{N}a_{-N}\widetilde{a}_{-N}
          \right] \ket{x}_T,\\
  \dket{\NN_T(x)} &=
     \exp\! \left[
       -\sum_{N = \frac{1}{2}, \frac{3}{2}, \dots}
        \frac{1}{N}a_{-N}\widetilde{a}_{-N}
          \right] \ket{x}_T,
\eea
where $\ket{x}_T \,(x = 0, \pi)$
are the ground states of the twisted sectors.
The overlaps of these states read
\bea
   \dbra{\DD_T(x)} \tilde{q}^{H_c} \dket{\DD_T(x')} &=
   \dbra{\NN_T(x)} \tilde{q}^{H_c} \dket{\NN_T(x')}  =
     \delta_{x,x'}
     \frac{1}{\eta({\tilde q})}
     \sum_{N \in \Z} \tilde{q}^{(N+\frac{1}{4})^2} , \\
   \dbra{\DD_T(x)} \tilde{q}^{H_c} \dket{\NN_T(x')}
   &=
     \delta_{x,x'} \frac{1}{\eta({\tilde q})}
     \sum_{N \in \Z} (-1)^N \tilde{q}^{(N + \frac{1}{4})^2} .
\eea

\bigskip
\section{Conformal boundary states in $S^1/\Z_2$}
\label{sec:conformal}
\subsection{Conventions}

We denote by $\dket{j;m,n}$ the Ishibashi states for
the degenerate representation $(j;m,n)$ of the Virasoro algebra
with $c=1$ \cite{GR}.
$\dket{j;m,n}$ is normalized as follows,
\beq
  \dbra{j;m,n} q^{H_c} \dket{j';m',n'} =
  \delta_{jj'} \delta_{mm'} \delta_{nn'} \chi_j(q), \quad
  \chi_j(q) = \frac{1}{\eta(q)} (q^{j^2} - q^{(j+1)^2}) .
\eeq
The $u(1)$ Ishibashi states $\dket{\DD(n)}, \dket{\NN(m)}$ for $r=1$ are
expressed as
\bea
  \dket{\DD(n)} &= \sum_{j \ge \frac{\abs{n}}{2}}
                  \dket{j; \tfrac{n}{2}, \tfrac{n}{2}} , \\
  \dket{\NN(m)} &= \sum_{j \ge \frac{\abs{m}}{2}} (-1)^{j-\tfrac{m}{2}}
                  \dket{j; \tfrac{m}{2}, -\tfrac{m}{2}} .
\eea
Since $S^1$ at $r=1$ is equivalent with the $SU(2)$ WZW model at level 1,
the conformal boundary states at $r=1$ are parametrized
by $g \in SU(2)$ \cite{GRW,GR},
\beq
  \ket{g} = 2^{-1/4} \sum_{j,m,n} D^{j}_{m,n}(g) \dket{j; m,n} .
\eeq
The overlap of these states reads
\beq
  \bra{g_1} \tilde{q}^{H_c} \ket{g_2} =
  \frac{1}{\eta(q)} \sum_{n \in \Z} q^{(n + \frac{\alpha}{2\pi})^2} ,
\label{eq:conformal_overlap}
\eeq
where $\alpha$ is determined by the trace in the fundamental representation
of $su(2)$,
\beq
  \trace_{j=\frac{1}{2}} (g_1^\dagger g_2) = 2 \cos \alpha .
\eeq
Among these conformal boundary states,
we can identify the Dirichlet and the Neumann states,
\bea
  \ket{\DD[\alpha]} &=
  2^{-1/4} \sum_{n \in \Z} e^{-in \alpha} \dket{\DD(n)} \notag\\
  &=
  2^{-1/4} \sum_{n \in \Z} e^{-in \alpha}
  \sum_{j \ge \frac{\abs{n}}{2}}
                  \dket{j; \tfrac{n}{2}, \tfrac{n}{2}} =
  \ket{g = \bigl(
  \begin{smallmatrix} e^{-i\alpha} & 0 \\ 0 & e^{i\alpha} \end{smallmatrix}
           \bigr) }, \\
  \ket{\NN[\beta]} &=
  2^{-1/4} \sum_{m \in \Z} e^{-im \beta} \dket{\NN(m)} \notag \\
  &=
  2^{-1/4} \sum_{m \in \Z} e^{-im \beta}
  \sum_{j \ge \frac{\abs{m}}{2}} (-1)^{j-\tfrac{m}{2}}
                  \dket{j; \tfrac{m}{2}, -\tfrac{m}{2}} =
  \ket{g = \bigl(
  \begin{smallmatrix} 0 & e^{-i\beta} \\ -e^{i\beta} & 0  \end{smallmatrix}
           \bigr) } .
\eea

\subsection{Conformal boundary states in $S^1(r=2)/\Z_2$}

We can construct the conformal boundary states of the orbifold
$S^1/\Z_2$ at the rational radii starting from those of $S^1$.
Here we report only the case of $r=2$. The case of $r \neq 2$
can be treated in the similar way \cite{Yamaguchi}.

The orbifold $S^1(r=2)/\Z_2$ can be considered as
the orbifold of the $SU(2)$ WZW model at level 1
by $\Gamma \subset SU(2)/\{1, -1\}$,
\beq
\Gamma = \{1, \gamma_1 = i\sigma_1, \gamma_2 = i\sigma_2,
              \gamma_3 = i\sigma_3 \}
         \cong D_2 = \Z_2 \times \Z_2 .
\eeq
Here $\gamma_3$ changes the radius of $S^1$ from 1 to 2,
while $\gamma_2$ does the reflection $X \mapsto -X$.
$\Gamma$ acts on the boundary states as
\beq
  \gamma : \, \ket{g} \mapsto \ket{\gamma g \gamma^{-1}} , \quad
  \gamma \in \Gamma .
\eeq
The generic boundary states in the orbifold theory can therefore be written
in the form
\beq
  \ket{g}_{\Gamma} = \frac{1}{\sqrt{\abs{\Gamma}}}
  \sum_{\gamma \in \Gamma} \ket{\gamma g \gamma^{-1}} , \quad
  g \in SU(2) ,
\label{eq:Gamma_state}
\eeq
unless $g$ belongs to the fixed points of $\Gamma$.

We have three types of fixed points
$F_a (a=1,2,3)$
corresponding to three subgroups $\{1, \gamma_a \} \cong \Z_2$
of $\Gamma$,
\beq
  F_a = \{g \in SU(2) |  \gamma_a g \gamma_a^{-1} =  g \} .
\eeq
We can parametrize $F_a$ as follows,
\bea
  F_1 &= \left\{ f_1(\tilde{\theta}) \equiv
            \begin{pmatrix} \cos \tilde{\theta} & i \sin \tilde{\theta} \\
                           i\sin \tilde{\theta} & \cos \tilde{\theta}
            \end{pmatrix} | \, 0 \le \tilde{\theta} < 2\pi \right\} , \\
  F_2 &= \left\{ f_2(\theta) \equiv
            \begin{pmatrix} \cos \theta & \sin \theta \\
                           -\sin \theta & \cos \theta
            \end{pmatrix} | \, 0 \le \theta < 2\pi \right\} , \\
  F_3 &= \left\{ f_3(\alpha) \equiv
            \begin{pmatrix} e^{i\alpha} & 0 \\
                            0 & e^{-i\alpha}
            \end{pmatrix} | \, 0 \le \alpha  < 2\pi \right\} .
\eea
$F_a$'s intersect at $g=\pm 1$,
\beq
  F_1 \cap F_2 \cap F_3 = \{\pm 1 \} .
\eeq
Clearly, $g = \pm 1 \in SU(2)$ consist of the fixed point set of $\Gamma$.

In order to resolve these fixed points, we need additional Ishibashi states
from the twisted sectors.
We have three types of twisted sectors corresponding to three types
of fixed points $F_a (a = 1,2,3)$.
For the fixed points $F_3$, the necessary states are given by the
$u(1)$ Ishibashi states at $r=1$,
\beq
  \dket{\TT_3(n)} \equiv
  \frac{1}{\sqrt{2}}
  (\dket{\DD(n+\tfrac{1}{2})} + \dket{\DD(-n-\tfrac{1}{2})}) \quad
  (n \in \Z_{\ge 0} ) .
\label{eq:T3}
\eeq
The Dirichlet Ishibashi states in this equation coincide
with those for $r=2$,
since their momenta take values in half integers in the unit of $r=1$
(see eq.\eqref{eq:pspec}).
Then one can resolve $F_3$ to yield the fractional states
\bes
  \ket{f_3(\alpha)}_\Gamma
  &= \frac{1}{2}(\ket{f_3(\alpha)} + \ket{f_3(-\alpha)})
  + 2^{1/4} \dket{\TT_3[\alpha]} ,  \\
  \dket{\TT_3[\alpha]} &= \sum_{n \in \Z_{\ge 0}}
  \cos((n + \tfrac{1}{2})\alpha) \dket{\TT_3(n)}  \quad
  (0 \le \alpha < 2\pi) .
\label{eq:f3}
\ees
One can easily show that
these resolved states are nothing but
the usual (non-fractional) $D0$-branes in $S^1/\Z_2$ at $r=2$,
\beq
  \ket{f_3(\alpha)}_\Gamma = \ket{\DD[\alpha/2]}_\mathcal{O} .
\eeq

For $F_1$ and $F_2$, the relevant twisted sectors are those coming from
the fixed points $X = 0, \pi$ of the reflection $X \mapsto -X$.
Within the chiral algebra
$u(1)/\Z_2$, we have four Ishibashi states from these twisted
sectors, namely,
$\dket{\DD_T(x)}$ and $\dket{\NN_T(x)}\, (x= 0, \pi)$.
These states are decomposed into infinite number of the Virasoro
Ishibashi states.
Actually, the overlaps of these states with themselves are
decomposed into the Virasoro characters as follows,
\bea
  \dbra{\DD_T(x)} q^{H_c} \dket{\DD_T(x)} &=
  \frac{1}{\eta(q)} \sum_{n \in \Z} q^{(n + \frac{1}{4})^2} =
  \sum_{n \in \Z} \chi_{\abs{n + \frac{1}{4}}} , \\
  \dbra{\DD_T(x)} q^{H_c} \dket{\NN_T(x)} &=
  \frac{1}{\eta(q)} \sum_{n \in \Z} (-1)^n q^{(n + \frac{1}{4})^2} =
  \sum_{n \in \Z} (-1)^n \chi_{\abs{n + \frac{1}{4}}} .
\eea
Consequently, the Ishibashi states
$\dket{\DD_T(x)}, \dket{\NN_T(x)}$ can be written in the form
\bea
  \dket{\DD_T(x)} &= \sum_{n \in \Z} \dket{\abs{n+\tfrac{1}{4}}; x} , \\
  \dket{\NN_T(x)} &= \sum_{n \in \Z} (-1)^n \dket{\abs{n+\tfrac{1}{4}}; x} ,
\eea
where $\dket{j; x}$ is the Virasoro Ishibashi state
with dimension $j^2$ from the twisted sector of $X = x$.
Using these states, we can write down the Ishibashi states
for the resolution of the fixed points $F_1$ and $F_2$,
\bea
  \dket{\TT_1(n)} &\equiv \frac{1}{\sqrt{2}}
  ( \dket{\abs{\tfrac{n}{2} + \tfrac{1}{4}}; 0} -
    \dket{\abs{\tfrac{n}{2} + \tfrac{1}{4}}; \pi} ) ,
  \label{eq:T1} \\
  \dket{\TT_2(n)} &\equiv \frac{1}{\sqrt{2}}
  ( \dket{\abs{\tfrac{n}{2} + \tfrac{1}{4}}; 0} +
    \dket{\abs{\tfrac{n}{2} + \tfrac{1}{4}}; \pi} ) \quad
  (n \in \Z_{\ge 0}) .
  \label{eq:T2}
\eea
These Ishibashi states $\dket{\TT_a(n)} (a = 1,2,3; n \in \Z_{\ge 0})$
from the twisted sectors satisfy
\beq
  \dbra{\TT_a(m)} q^{H_c} \dket{\TT_b(n)} =
  \delta_{ab} \delta_{mn} \frac{1}{\eta(q)}
  q^{\frac{1}{4}(n + \frac{1}{2})^2} .
\eeq
Then the resolved boundary states can be constructed in a way parallel
to eq.\eqref{eq:f3}, namely,
\bes
  \ket{f_a(\beta)}_\Gamma &=
  \frac{1}{2}(\ket{f_a(\beta)} + \ket{f_a(-\beta)}) +
  2^{1/4} \dket{\TT_a[\beta]} ,\\
  \dket{\TT_a[\beta]} &= \sum_{n \in \Z_{\ge 0}}
  \cos((n + \tfrac{1}{2})\beta) \dket{\TT_a(n)} \quad
  (0 \le \beta < 2\pi) .
\label{eq:fa}
\ees
Among these states, we can find the fractional $D1$-branes
of $S^1(r=2)/\Z_2$,
\bes
  \ket{f_1(\pi/2)}_\Gamma &= \ket{\NN[\pi]; +}_\mathcal{O} , \quad
  \ket{f_1(3\pi/2)}_\Gamma = \ket{\NN[\pi]; -}_\mathcal{O} , \\
  \ket{f_2(\pi/2)}_\Gamma &= \ket{\NN[0]; +}_\mathcal{O} , \quad
  \ket{f_2(3\pi/2)}_\Gamma = \ket{\NN[0]; -}_\mathcal{O} .
\ees

At $g=\pm 1$, we might expect further resolution
since all the elements $\gamma \in \Gamma$ fix $g=\pm 1$.
For $g=1$, we actually need one more resolution
and the result is given by the fractional $D0$-branes of the orbifold
\bes
  \ket{\DD[0]; \pm}_\mathcal{O}
  &= \frac{1}{2}\ket{g=1} + 2^{1/4} \frac{1}{2}
  \Bigl(\dket{\TT_3[0]} \pm (\dket{\TT_2[0]} + \dket{\TT_1[0]}) \Bigr) , \\
  \ket{\DD[\pi]; \pm}_\mathcal{O}
  &= \frac{1}{2}\ket{g=1} + 2^{1/4} \frac{1}{2}
  \Bigl(-\dket{\TT_3[0]} \pm (\dket{\TT_2[0]} + \dket{\TT_1[0]}) \Bigr) .
\ees
For $g=-1$, however, all the fractional branes in \eqref{eq:fa}
give the same result
\beq
  \ket{f_1(\pi)}_\Gamma = \ket{f_3(\pi)}_\Gamma = \ket{f_3(\pi)}_\Gamma =
  \ket{g = -1} .
\eeq
Clearly, this state can not be decomposed any more and
we have no resolution at $g=-1$.



\end{document}